\documentclass[useAMS,usenatbib]{mn2e}

\usepackage{graphicx}
\usepackage{times}
\usepackage{natbib}
\usepackage{amsmath}

\newcommand{\mone}{$^{-1}$}

\newcommand{\ypar}{$y$-parameter}

\newcommand{\lcdm}{$\Lambda$CDM}
\newcommand{\omegam}{$\Omega_{\rm m}$}

\newcommand{\omegab}{$\Omega_{\rm b}$}

\newcommand{\gadgettwo}{\textsc{gadget-2}}
\newcommand{\gadgetthree}{\textsc{gadget-3}}
\newcommand{\msun}{M$_\odot$}

\newcommand{\rtwoh}{$R_{200}$}
\newcommand{\hmone}{$\,h^{-1}$}

\newcommand{\planck}{{\it Planck}}
\newcommand{\ovisc}{{\it NR}}
\newcommand{\csfw}{{\it CSF}}
\newcommand{\csfwbh}{{\it AGN}}

\newcommand{\chandra}{{\it Chandra}}
\newcommand{\suzaku}{{\it Suzaku}}

\newcommand{\crho}{$C$}
\newcommand{\crhov}{$C_\mathcal{R}$}
\newcommand{\crhoest}{$C_\mathcal{R}^{\rm est}$}
\newcommand{\sigmasb}{$\sigma_{\rm SB}$}
\newcommand{\sigmay}{$\sigma_y$}
\newcommand{\mgas}{$M_{\rm gas}$}
\newcommand{\mhe}{$M_{\rm he}$}
\newcommand{\fgas}{$f_{\rm gas}$}
\newcommand{\pmone}{$^{-1}$}
\newcommand{\pmtwo}{$^{-2}$}

\def\aap{A\&A}
\def\apj{ApJ}

\def\apjl{ApJ}
\def\mnras{MNRAS}
\def\araa{ARA\&A}

\def\physrep{Phys. Rep.}

\def\apss{Ap\&SS}      
\def\apjs{ApJS}
\def\pasj{PASJ}

\def\pasp{PASP}

\title[Large-scale ICM inhomogeneities: improving mass estimates]
{Large-scale inhomogeneities of the intracluster medium: improving mass estimates using 
the observed azimuthal scatter}
\author[M. Roncarelli et al.]
{M.~Roncarelli$^{1,2}$, 
S.~Ettori$^{2,3}$, 
S.~Borgani$^{4,5,6}$,
K.~Dolag$^{7,8}$,
D.~Fabjan$^{5,6,9,10}$ and \newauthor 
L.~Moscardini$^{1,2,3}$ \newauthor
 \\~\\ 
$^1$Universit\`a di Bologna, Dipartimento di Fisica e Astronomia, viale Berti Pichat 6/2, 
I-40127 Bologna, Italy \\
$^2$Istituto Nazionale di Astrofisica (INAF) - Osservatorio Astronomico di Bologna, via 
Ranzani 1, I-40127 Bologna, Italy \\
$^3$Istituto Nazionale di Fisica Nucleare (INFN) - Sezione di Bologna, viale Berti Pichat 
6/2, I-40127 Bologna, Italy \\
$^4$Universit\`a di Trieste, Dipartimento di Fisica, Sezione di Astronomia, via Tiepolo 
11, I-34131 Trieste, Italy \\
$^5$Istituto Nazionale di Astrofisica (INAF) - Osservatorio Astronomico di Trieste, via 
Tiepolo 11, I-34131 Trieste, Italy \\
$^6$Istituto Nazionale di Fisica Nucleare (INFN) - Sezione di Trieste, Via Valerio 2, 
I-34127 Trieste, Italy \\
$^7$Universit\"atssterenwarte M\"unchen, M\"unchen, Germany \\
$^8$Max-Planck-Institut f\"ur Astrophysik, Garching, Germany \\
$^9$SPACE-SI, Slovenian Centre of Excellence for Space Sciences and Technologies, 
A\u{s}ker\u{c}eva 12, 1000 Ljubljana, Slovenia\\
$^{10}$University of Ljubljana, Faculty of Mathematics and Physics, Jadranska 19, 1000 
Ljubljana, Slovenia
}
\begin{document}

\pagerange{\pageref{firstpage}--\pageref{lastpage}} \pubyear{2012}

\maketitle

\label{firstpage}

\begin{abstract}
Achieving a robust determination of the gas density profile in cluster outskirts is a 
crucial point in order to measure their baryonic content and to use them as 
cosmological probes. The difficulty in obtaining this measurement lies not only in the 
low surface brightness of the ICM, but also in the inhomogeneities of the gas associated 
to clumps, asymmetries and accretion patterns.
Using a set of hydrodynamical simulations of 62 galaxy clusters and groups 
we study this kind of inhomogeneities, focusing on the ones on the large scale that, 
unlike clumps, are the most difficult to identify. To this purpose we introduce the 
concept of \emph{residual clumpiness}, \crhov, that quantifies the 
large-scale inhomogeneity of the ICM. After showing that this quantity can be robustly 
defined for relaxed systems, we characterize how it varies with radius, mass 
and dynamical state of the halo. Most importantly, we observe that it introduces an 
overestimate in the determination of the density profile from the X-ray emission, which 
translates into a systematic overestimate of 6 (12) per cent in the measurement of \mgas\ 
at \rtwoh\ for our relaxed (perturbed) cluster sample. At the same time, the increase of 
\crhov\ with radius introduces also a $\sim$2 per cent systematic underestimate in the 
measurement of the hydrostatic-equilibrium mass (\mhe), which adds to the previous one 
generating a systematic $\sim$8.5 per cent overestimate in \fgas\ in our relaxed sample. 
Since the residual clumpiness of the 
ICM is not directly observable, we study its correlation with the azimuthal scatter in 
the X-ray surface brightness of the halo, a quantity that is well-constrained by current 
measurements, and in the \ypar\ profiles that is at reach of the forthcoming SZ 
experiments. We find that their correlation is highly significant ($r_{\rm S} = 
0.6-0.7$), allowing to define the azimuthal scatter measured in the X-ray surface 
brightness profile and in the \ypar\ as robust proxies of \crhov. After providing a 
function that connects the two quantities, we obtain that correcting the observed 
gas density profiles using the azimuthal scatter eliminates the bias in the measurement 
of \mgas\ for relaxed objects, which becomes $0 \pm 2$ per cent up to 2\rtwoh, and 
reduces it by a factor of 3 for perturbed ones. This method allows also to eliminate the 
systematics on the measurements of \mhe\ and \fgas, although a significant halo to halo 
scatter remains.
\end{abstract}



\section{Introduction} \label{sec:intro}
Clusters of galaxies are the most massive virialized systems of the Universe, they form 
in the knots of the cosmic web from which they continuously accrete material in the form 
of dark matter, gas and galaxies. Their importance for astrophysics is crucial, since 
they enclose information on the large-scale structure formation \citep[see][for a 
review]{kravtsov12} and provide constraints on cosmological parameters which are 
independent with respect to cosmic microwave background (CMB), type-Ia supernovae and 
galaxy surveys \citep[see, e.g.,][and references therein]{allen11}.

The precision with which clusters can be used as cosmological probes depends on the 
accuracy in the measurements of their mass, not only in terms of the total mass 
but also for its baryonic fraction $f_{\rm b} \equiv M_{\rm b}/M_{\rm tot}$. If 
the value of $f_{\rm b}$ measured inside clusters matches the cosmic one 
(\omegab/\omegam) it can be used to obtain an independent estimate of \omegam. However we 
know that the assumption of $f_{\rm b} \simeq \Omega_{\rm b}/\Omega_{\rm m}$ holds only 
at large radii thus requiring the achievement of accurate observations also of cluster 
outskirts. In the recent years this point raised a great interest of the astrophysical 
community towards the study of clusters around and beyond the virial radius.

When moving outside the core the intracluster medium (ICM) is often affected by the 
presence of inhomogeneities and substructures, whose impact on the observed properties 
has been studied with a statistical approach \citep[see, e.g.,][]{jeltema05,
bohringer10,andrade-santos12}. The difficulty in observing cluster outskirts lies in the 
very low X-ray surface brightness of the ICM, that drops below the diffuse 
extragalactic X-ray background at $r\simeq$\rtwoh\footnote{In this work \rtwoh\ is defined as 
the radius enclosing an average density equal to 200 times the critical density of the 
Universe $\rho_{\rm c} \equiv \frac{3 H_0}{8\pi G}$.} \citep[see][]{roncarelli06a,
roncarelli06b}. A possible method consists in stacking the images of different objects, 
as done by \cite{eckert12} who measured the density profiles and the azimuthal 
scatter in the X-ray surface brightness of 31 ROSAT-PSPC objects, showing a 
clear segregation between Cool-Core (CC) and No-Cool-Core (NCC) systems over the radial 
range 0.1--0.8 $R_{200}$, with the first ones exhibiting a scatter of 20–-30 per cent, 
which corresponds to density variations of the order of 10 per cent. Above 0.8 \rtwoh, 
the azimuthal scatter increases up to values of 60--80 per cent both in CC and NCC 
clusters, suggesting that they experience common physical conditions in shaping their 
X-ray surface brightness profiles.
However the limit of the stacking technique is that it does not allow a precise 
determination of the temperature profiles of single objects. In this framework a great 
step forward has come from the \suzaku\ X-ray satellite that, thanks to its very low 
instrumental background, provided the first 
robust spectroscopical analyses of the regions close to \rtwoh\ for the 
brightest objects \citep[see, e.g.,][]{george09,bautz09,hoshino10,kawaharada10,
simionescu11,humphrey12,sato12}. These works 
also highlighted the presence of a substantial 
azimuthal scatter, likely associated to clumps or inhomogeneities on the large scales, 
induced by major or even minor mergers that can cause sloshing and swirling motions in 
the ICM \citep[see also][]{sanders12,churazov12,simionescu12}.

A promising complementary view with respect to X-rays can come from the observations of 
the thermal Sunyaev-Zel’dovich effect \citep[SZ,][]{sunyaev70,birkinshaw99,rephaeli05}. 
The SZ signal is associated to temperature fluctuations in the CMB 
spectrum that are directly proportional to the integral of the electron pressure along 
the line of sight. The SZ effect, in combination with the X-ray emission, can therefore 
probe the triaxial structure of the gas \citep{defilippis05,morandi12,sereno12}. 
Moreover, if the source is sufficiently extended to be resolved spatially even with 
the large beams available with present instruments such as \planck\ \citep[as for nearby, 
hot systems like Coma,][]{planck12} or multi-pixel bolometer arrays (as the 
\emph{APEX-SZ} experiment, \citeauthor{basu10} \citeyear{basu10}, or \emph{Bolocam} at 
the Caltech Submillimeter Observatory, \citeauthor{sayers12} \citeyear{sayers12}) it can 
be used also to directly infer the pressure profile, under the assumption of the 
spherical symmetry of the ICM distribution, out to a significant fraction of the virial 
radius \citep[see also][]{walker12,eckert13}.

From the theoretical point of view the modelisation of cluster outskirts is 
affected by different uncertainties with respect to cluster cores. On one side 
different feedback mechanisms do not affect significantly the behavior of temperature and 
density profiles, with gravity that constitutes the dominating physical process 
\citep{roncarelli06b}. On the other hand, the presence of shocks and turbulence may 
lead to the break of the hydrostatic equilibrium \citep[see, e.g.,][and references 
therein]{iapichino08,vazza09,burns10,nagai11}.
In addition to that, hydrodynamical simulations showed how the 
outer ICM is affected by the presence of clumps \citep{roncarelli06b,nagai11} and 
inhomogeneities \citep{kawahara08,vazza11} that may bias the reconstruction of the 
clusters' properties and explain the observed azimuthal scatter \citep[see also][]
{vazza12,zhuravleva13}. These large-scale inhomogeneities, both in gas densities and 
temperatures, are responsible for half of the $\sim 30$ per cent underestimate in the 
X-ray reconstructed hydrostatic mass, as highlighted by numerical simulations \citep[see, 
e.g.,][]{rasia12,piffaretti08}. The remaining part is mostly due to the residual 
bulk motions of the ICM, that preserve energy in non-thermalised form, not 
traceable with the measures of the density and temperature profiles alone \citep[see, 
e.g.,][and references therein]{rasia06,battaglia12,nelson12}, but potentially 
detectable with spatially-resolved X-ray spectroscopy (see, e.g., the 
works on the Coma clusters by \citeauthor{schuecker04} \citeyear{schuecker04} and
\citeauthor{churazov12} \citeyear{churazov12}).

In the work presented here we study the gas inhomogeneities present in the 
outskirts (close to \rtwoh\ and beyond) of galaxy clusters and groups using a set of 
hydrodynamical simulations, with the objective of analyzing how the 
different physical processes may affect the degree of density fluctuations, 
together with the halo mass and its dynamical status. We concentrate our efforts in the 
definition of the density inhomogeneities associated to large-scale fluctuations, 
for which we introduce the concept of \emph{residual clumpiness}, i.~e. the clumpiness 
of the ICM after obvious clumpy structures have been removed. We discuss how this 
phenomenon may bias high the X-ray density profile measurements and explore, for the 
first time, the possibility of obtaining a direct measurement of this quantity via the 
analysis of the observed azimuthal scatter in the X-ray and \ypar\ profiles of single 
objects. This method of estimating the residual clumpiness of the ICM allows us to 
propose a technique to reduce this bias and, therefore, to improve the precision of the 
measurement of the total mass and of the gas mass of galaxy clusters.

This paper is organized as follows. In the next Section we describe the set of 
hydrodynamical simulations used for our study. In Section~\ref{sec:crho} we characterize 
the gas inhomogeneities of our simulated haloes, we provide the definition of the 
residual clumpiness (\crhov) and we study its dependence on cluster properties. 
Section~\ref{sec:scatter} presents our results on the correlation between the azimuthal 
scatter and large-scale inhomogeneities of cluster outskirts. In 
Section~\ref{sec:corr_mass} we quantify how the residual clumpiness biases the 
measurements of the gas mass, the hydrostatic-equilibrium mass and the gas fraction of 
the haloes and present a method to correct them using the observed azimuthal scatter. We 
summarize and draw our conclusions in Section~\ref{sec:concl}.


\section{The simulated clusters} \label{sec:sims}

The clusters object of our work belong to a set of 29 high-resolution re-simulations of 
galaxy cluster regions. A detailed description of the whole procedure to obtain them 
can be found in \cite{bonafede11}. Here we provide a summary.

\subsection{Simulation parameters}

The cosmology assumed is a flat \lcdm\ model with $\Omega_{\rm m}=0.24$ and 
$\Omega_{\Lambda} \equiv 1-\Omega_{\rm m}=0.76$, a Hubble parameter $h=0.72$ (being the 
Hubble constant $H_0=100 \, h$ km s$^{-1}$ Mpc$^{-1}$); we fix the primordial power 
spectrum of the DM fluctuations with slope $n=0.96$ and normalization coherent with 
$\sigma_8=0.8$.

With these assumptions we carried out a large dark-matter (DM) only simulation using the 
\gadgetthree\ code, an evolved version of \gadgettwo\ \citep{springel05}, with a periodic 
box of 1\hmone\ Gpc on a side and then identified the clusters in the $z=0$ output with a 
\emph{Friends of Friends} (FOF) algorithm, with linking length fixed to 0.17 times the 
mean inter-particle separation. The Lagrangian regions around $7R_{\rm vir}$ from the 
centre of the 24 most massive objects (all with $M_{\rm vir} > 10^{15} h^{-1}$ \msun) 
were identified, traced back to their initial positions and then re-simulated at higher 
resolution using the zoomed initial conditions technique \citep{tormen97}. These 
re-simulations were run adding the hydrodynamical part by turning on the \emph{Smoothed 
Particles Hydrodynamics} (SPH) code implemented in \gadgettwo. This was done assuming 
$\Omega_{\rm b}=0.045$, with mass resolutions of $8.4 \times 10^8$\hmone \msun\ and $1.6 \times 
10^8$\hmone \msun\ for the DM and gas particles, respectively, and with a 
Plummer-equivalent softening length for the gravitational force fixed to $\varepsilon=5 
\, h^{-1}$ kpc in physical units at $z<2$ and kept fixed in comoving units at earlier 
times.

In order to add statistics at the lower mass scales, another set of 5 galaxy clusters 
with masses $M_{200} = 1-5 \times 10^{14}$\hmone\msun\ was selected at $z=0$ and 
re-simulated with the same method, reaching a total of 29 re-simulated regions.


\subsection{Feedback models}
\label{ssec:feedback}

The set of 29 re-simulations was run with three different physical implementations. More 
detailed descriptions of these physical models can be found in \cite{planelles12}.

\begin{itemize}
\item[{\it (i)}] \ovisc: non-radiative runs. They do not include any physical process 
except gravitation and hydrodynamics. Although unrealistic, they are useful as a test to 
check the impact of the various physical mechanisms.

\item[{\it (ii)}] \csfw: runs including cooling, star formation, metal enrichment and 
galactic winds. The radiative cooling rates are computed following the procedure of 
\cite{wiersma09}, considering also the contribution of metals in the hypothesis of a 
gas in (photo-)ionisation equilibrium by using the \textsc{cloudy} code 
\citep{ferland98}.
Star formation is followed according to the multi-phase model 
of \cite{springel03} which also includes energy release by supernovae (SNe). Other 
feedback sources include heating from a spatially uniform, time dependent UV background 
\citep{haardt96} and metals released by type-II and type-Ia SNe and AGB stars 
according to the model of \cite{tornatore07}. Finally, a kinetic feedback associated to 
galactic ejecta is implemented assuming a mass upload proportional to the star formation 
rate and a wind speed of $v_{\rm w}=500$ km s\mone.

\item[{\it (iii)}] \csfwbh: like \csfw, but with wind speed reduced to 
$v_{\rm w}=350$ km s\mone\ and including also the feedback associated to gas 
accretion onto supermassive black holes (BHs). Its implementation is similar to the 
one of \cite{springel05}, with the inflow of matter that proceeds at Bondi rate up to the 
Eddington limit. In our implementation SPH particles stochastically selected for 
providing accretion into the BH feed it only with 1/4 of their mass at time, thus 
mimicking a more continuous flow of material. The amount of kinetic energy released 
by each BH particle is given by 
\begin{equation}
\dot{E}_{\rm feed} = \epsilon_{\rm r}\epsilon_{\rm f} \dot{M}_{\rm BH}^2 \ ,
\end{equation}
being $\epsilon_{\rm r}$ and $\epsilon_{\rm f}$ the radiative efficiency and the 
fraction of energy coupled to the gas, respectively. In our implementation these two 
free parameters were fixed at $\epsilon_{\rm r}=0.1$ and $\epsilon_{\rm f}=0.05$, 
increasing to $\epsilon_{\rm f}=0.2$ when the accretion rate is smaller than 
one-hundredth of the Eddington limit \citep[see also][]{sijacki07,fabjan10}.
We will take this model as our reference one.
\end{itemize}


\subsection{Definition of the halo sample}
\label{ssec:sample}
\begin{figure*}
\includegraphics[width=0.85\textwidth]{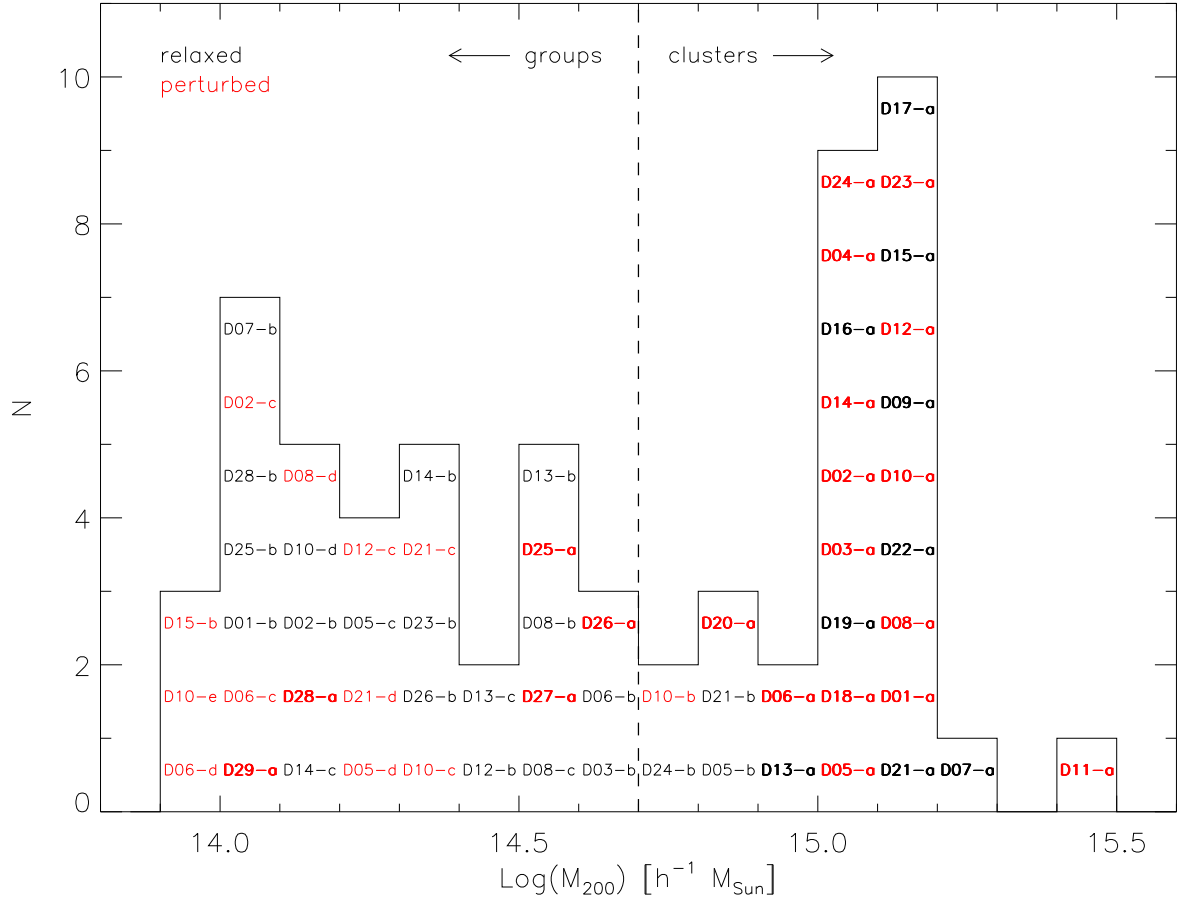}
\vspace{1cm}
\caption{Distribution of $M_{200}$ for the sample of 62 simulated haloes in our 
reference (\csfwbh) run. Cluster names correspond to the ones given in \protect 
\cite{bonafede11}, with the addition of D25--D29 that indicate the low mass 
objects re-simulations. The main haloes of each re-simulation are dubbed \emph{DXX--a} 
(bold face) while their satellites are indicated with letters from \emph{--b} to 
\emph{--e} in decreasing mass order. The vertical dashed line indicates the mass limit of 
$5 \times 10^{14}$\hmone\msun\ used to separate groups (34 objects) and clusters (28). 
Objects marked in red correspond to perturbed (32) haloes, while black indicates 
relaxed (30) ones, as defined in Section~\ref{ssec:crho_samp}.}
\label{fig:histo}
\end{figure*}

Since the volume of each of the 29 re-simulations encloses a spherical region of 
radius much larger than the virial radius of the main object, it is common to find other 
haloes included in the final $z=0$ outputs. Hence, in order to increase the statistical 
significance of our results and to characterize better also the low mass objects, we 
included in our sample all the `secondary' haloes with $M_{200} > 10^{14}$\hmone\msun\ 
in \emph{at least} one of the three runs. 
The values of $M_{200}$ were computed starting from the FOF catalogues and applying a 
\emph{Spherical Overdensity} algorithm. This increases our halo 
sample to a total of 62 objects in different mass ranges. A histogram of their masses is 
shown in Fig.~\ref{fig:histo}.

In order to study the dependence of the clumpiness on the mass of the halo and on its 
dynamical properties, we divide our objects into subsamples: given the amplitude of 
our initial sample, this can be done without losing statistical robustness. 
In particular we will adopt the following two definitions:
\begin{itemize}
\item[-] \emph{clusters}/\emph{groups} (28 and 34 objects, respectively), according to 
their $M_{200}$ being higher/lower than $5 \times 10^{14}$\hmone\msun,
\item[-] \emph{relaxed}/\emph{perturbed} (30 and 32 objects, respectively), according to 
a definition based on the clumpiness and described in Section~\ref{ssec:crho_samp}.
\end{itemize}

The histogram of Fig.~\ref{fig:histo} shows also these classifications.
It is worth to mention that clusters tend to be slightly more perturbed (16 and 12 
relaxed) than groups (16 and 18) due to their more active dynamical state.


\section{Density inhomogeneities} \label{sec:crho}

\subsection{The gas clumpiness}
A commonly used indicator of the degree of inhomogeneity of a medium is its clumpiness. 
Even if it is normally used to measure the amount of small clumps in an approximately 
uniform medium, its general definition allows us to use it also to quantify the amount 
of inhomogeneity associated to large-scale density fluctuations.

The clumpiness or clumping factor \crho\ of a fluid element is defined by the following 
formula:
\begin{equation}
C \equiv \frac{<\rho^2>_V}{<\rho>_V^2} \ ,
\label{eq:clp0}
\end{equation}
where $\rho$ is the fluid density and the brackets $<>_V$ indicate the average calculated 
over its volume. The clumpiness is therefore defined as being equal to unity for a 
perfectly uniform medium and $>1$ otherwise, with higher values indicating higher levels 
of inhomogeneities.
When writing explicitly the integrals, eq.~(\ref{eq:clp0}) can be expressed as as
\begin{equation}
C = \frac{V}{M^2} \int_V \rho^2 dV \ ,
\label{eq:clp1}
\end{equation}
where $M$ and $V$ are the mass and volume of the gas element, respectively.
Since we are using SPH simulations, for a given distance $r$ from the cluster centre, we 
compute $C(r)$ by converting eq.~(\ref{eq:clp1}) into
\begin{equation}
C(r) = \frac{\sum_i m_i \rho_i}{(\sum_i m_i)^2} \, V_{\rm shell} \ ,
\label{eq:clp2}
\end{equation}
where the sum $\sum_i$ is computed over the particles with distance between $r$ and 
$r+dr$, $m_i$ and $\rho_i$ are the mass and density, respectively, of the $i$-th SPH 
particle and $V_{\rm shell}=\frac{4}{3}\pi [(r+dr)^3 - r^3]$ is the volume of the shell 
used to compute the quantity.

For the purpose of our work, for every halo we compute the clumpiness and the other 
physical quantities in the range $0<r<2R_{200}$, in 50 equally spaced bins.


\subsection{Separating small clumps from large-scale inhomogeneities: the residual 
            clumpiness}
\label{ssec:volume}
The amount of clumpiness of the ICM is caused by two different phenomena: the presence of 
small dense clumps and the density fluctuations on larger scales associated to 
asymmetries in the large-scale accretion pattern, with the first one that constitutes the 
dominating contribution. Here we describe a method that proves useful to isolate these 
two components in order to treat them separately. This aspect is particularly important 
when we consider that the physical properties of clumps and their abundance depend on the 
processes of cooling and star formation and, consequently, their presence in our 
simulated clusters is subject to the uncertainties associated to the implementation of 
physical models in hydrodynamical codes. Moreover, several works \citep[see, e.g.,][and 
references therein]{mitchell09,sijacki12} highlighted how SPH simulations produce a 
higher quantity of dense clumps with respect to Eulerian ones. This is associated to the 
different numerical viscosity, intrinsic in the codes themselves, that causes a lower 
dissipation efficiency in SPH simulations.

To this purpose, we follow the volume-selection scheme described in \cite{roncarelli06b} 
to compute the profiles of galaxy clusters close to and beyond $R_{200}$. This method 
consists in sorting the SPH particles belonging to a given radial bin according to their 
physical density. Then, starting from the most diffuse one, we sum up their volumes 
(defined as $V_i = m_i/\rho_i$) until we reach a fixed fraction, 99 per cent in our case, 
of the total volume of the radial bin and consider only these particles to compute 
\crho: the remaining particles are identified as clumps and considered separately in 
our computation. This procedure proved to be effective in excising all the dense 
and cold clumps that are highly dependent on the physical assumptions, as well as the 
small X-ray bright regions that are usually masked out in observational analyses, 
thus providing a good description of the global properties of the ICM such as density, 
temperature and X-ray surface brightness \citep[see the discussion in][]{roncarelli06b}. 
Recently, a similar approach has also been applied to Eulerian simulations by 
\cite{zhuravleva13} who approximate the density distribution of the gas in different 
radial bins with a lognormal distribution and mark as clumps the cells with density above 
values fixed in terms of the $\sigma$ of the distribution \citep[see also][]{khedekar12}. 
In fact, although applied in completely different numerical simulations (grid-based and 
SPH), the two methods are formally identical, with their threshold of $f_{cut}=3.5$ that 
matches our 99 per cent criterion.
For what concerns the possible biases caused by unresolved clumpy gas, we have verified 
that they have a small impact on our results. For more details about this point and about 
the clumps physical properties, we refer to the Appendix~\ref{app:clumps_obs}.
We use this method to define 
the \emph{residual clumpiness} associated to large-scale perturbations, \crhov, and to 
separate it from the total one: this allows us to note that the value of \crhov\ is 
about one order of magnitude lower with respect to \crho\ inside \rtwoh\ (see the 
discussion in Section~\ref{ssec:crho_samp}).

\begin{figure*}
\includegraphics[width=0.9\textwidth]{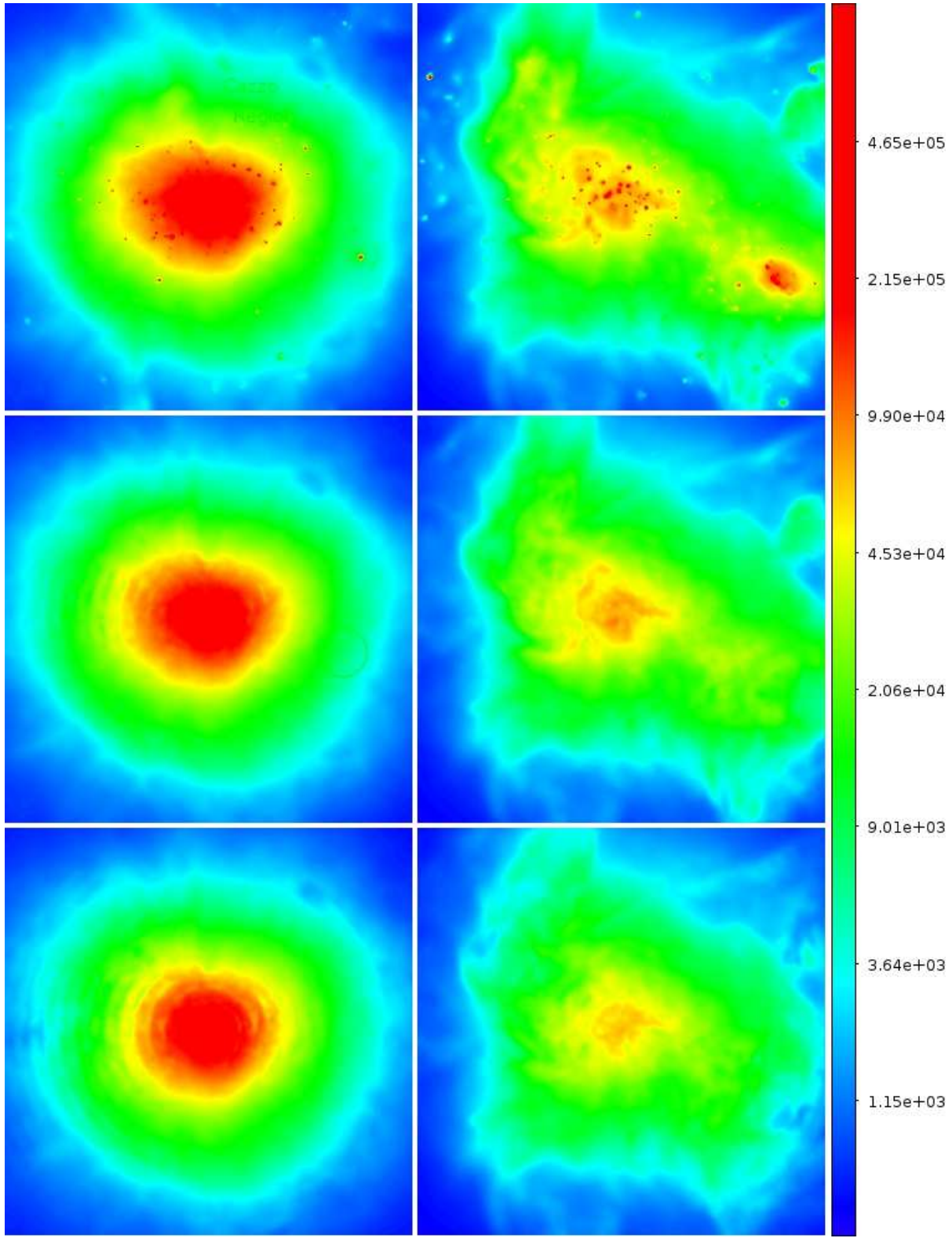}
\caption{Bolometric (0.1--10 keV) X-ray flux (in units of counts s\pmone\ 
cm\pmtwo) of a relaxed cluster (D17-a, left column) and a 
perturbed one (D12-a, right column). The size of each map is 1\rtwoh\ per side. In the 
first row all the particles have been considered, in the second and third ones the volume 
cut has been applied considering the 99 per cent and 95 per cent threshold, 
respectively.}
\label{fig:maps}
\end{figure*}

\begin{figure}
\includegraphics[width=0.49\textwidth]{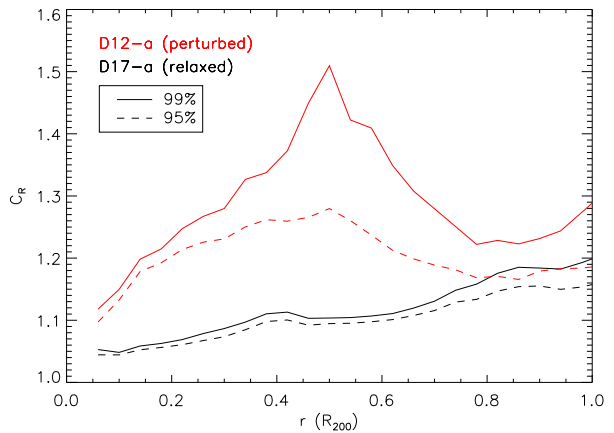}
\caption{Residual clumpiness as a function of distance from the centre computed 
for the two haloes shown in the maps of Fig.~\ref{fig:maps}: D17-a (black lines, lower 
values) and D12-\emph{a} (red lines, higher values). The value of \crhov\ has been 
computed adopting two different thresholds for the volume-clipping method: 99 per cent 
(solid lines) and 95 per cent (dashed lines). The physical model assumed is the reference 
one (\csfwbh). }
\label{fig:clp_99_95}
\end{figure}

We have also checked that in most of the cases lowering the threshold value to 95 
per cent (i.e. increasing by a factor of 5 the volume of gas considered as small-scale 
clumps) produces negligible changes in \crhov. However, when the cluster is perturbed by 
the presence of a large halo (e.g. an infalling group), lowering the density threshold 
for the clumps identification can change the final result by up to 50 per cent in the 
radial bins corresponding to the halo distance. A sketch of the application of this 
method is shown in Fig.~\ref{fig:maps} for two different cases. These X-ray surface 
brightness maps show that when cutting the 1 per cent densest volume (from top to middle 
panels) we are able to remove all obvious clumpy structures that are associated to the 
brightest peaks. When cutting the 5 per cent of the volume (bottom panels) the surface 
brightness remains almost unchanged for a relaxed halo (left) while for a dynamically 
perturbed one (right) the infalling halo on the bottom right is progressively removed, 
indicating that it is difficult to define a precise threshold to separate the two 
components. The effect on the definition of \crhov\ can be clearly seen in the plot of 
Fig.~\ref{fig:clp_99_95}, where we show the residual clumpiness as a function of radius 
computed for these two haloes by adopting the two different volume threshold: while for 
D17-\emph{a} the value of \crhov\ is almost independent on the threshold chosen up to 
\rtwoh, for D12-\emph{a} the presence of the infalling halo at $r \simeq 0.5$\rtwoh\ 
produces a difference of about 0.25.

For this reason we use the difference between the value of \crhov\ computed with these 
two limits to introduce another useful halo classification. We sort our haloes according 
to the maximum relative difference between $C^{99}(r)$ and $C^{95}(r)$ for 
$r<R_{200}$ and split our 62 haloes roughly into two equal subsamples: we define a halo 
as \emph{relaxed} when this difference is less than 8 per cent, and \emph{perturbed} 
otherwise. In this way, we end up with 32 perturbed haloes and 30 
relaxed ones. For this last set we can consider our value of \crhov\ to represent a 
robust estimate of the amount of inhomogeneities associated to large-scale asymmetries. 
We also verified that our classification has a good correspondence with an 
observational-like classification based on X-ray images. We refer the reader to 
Appendix~\ref{app:rel_ptb} for the details on this point.

\subsection{Dependence on physics and environment}
\label{ssec:crho_samp}

\begin{figure}
\includegraphics[width=0.49\textwidth]{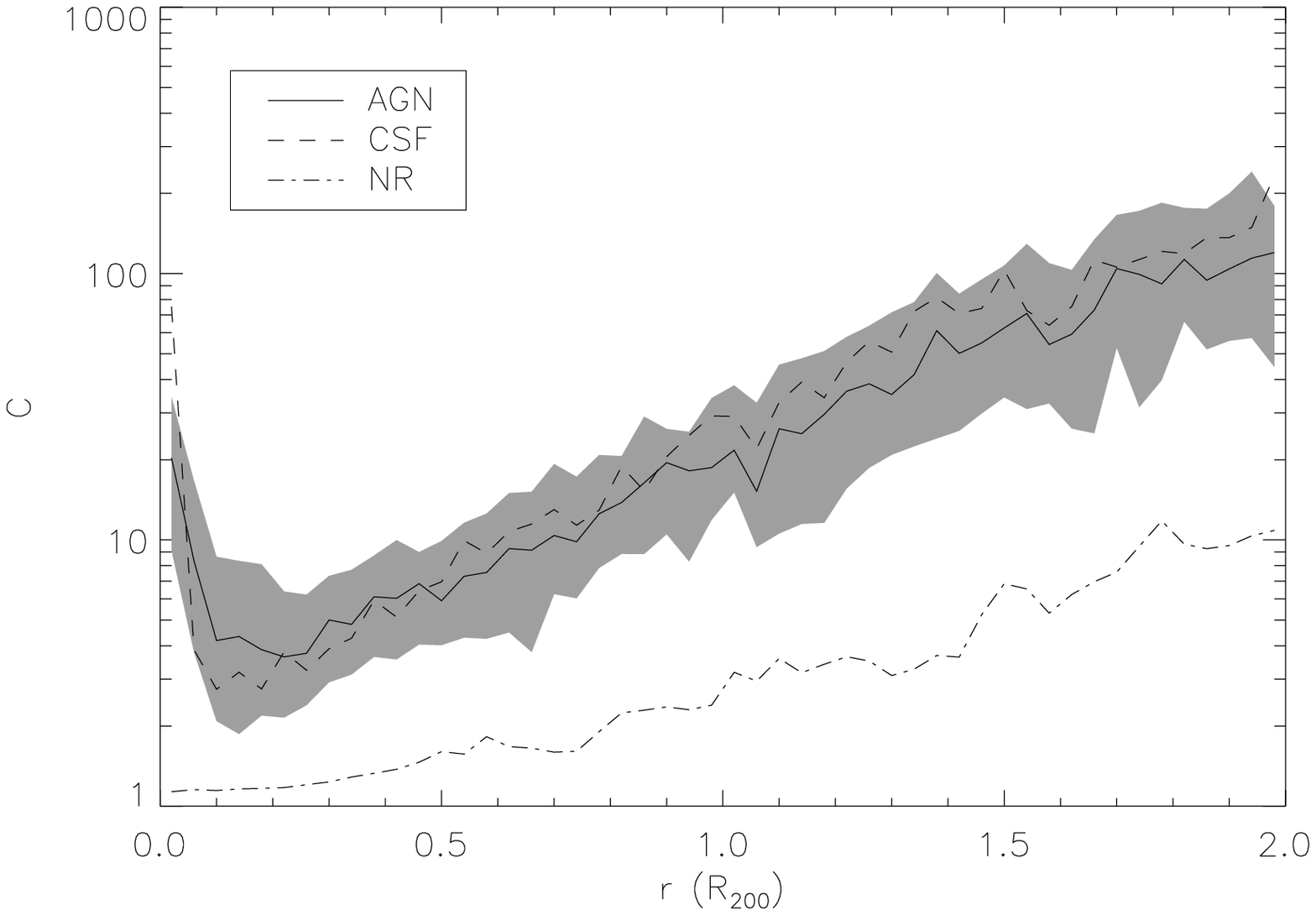}
\includegraphics[width=0.49\textwidth]{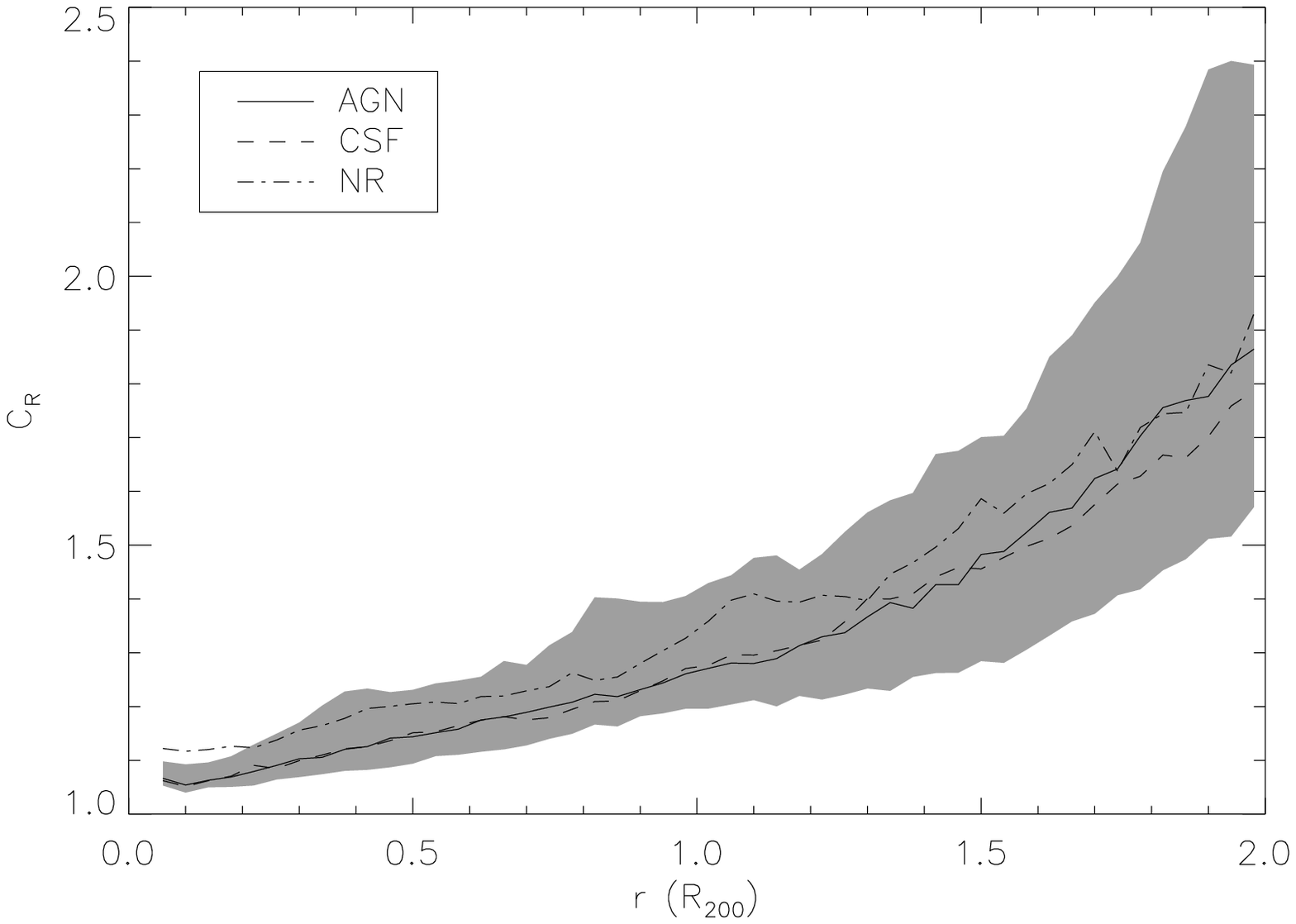}
\caption{Upper panel: gas clumpiness of our simulated haloes as a function of  
distance from the centre, for the \csfwbh\ (solid line), \csfw\ (dashed) and the \ovisc\ 
model (dot-dashed). The values represent the median of 62 objects, the grey-shaded region 
encloses the quartiles of the \csfwbh\ model (the quartile regions of the other two 
runs have similar sizes). Lower panel: same as upper panel but for the residual 
clumpiness, i.e. after applying our volume-clipping method.}
\label{fig:clp_phys}
\end{figure}

We show on the top panel of Fig.~\ref{fig:clp_phys} the results of the clumpiness as a 
function of radius for our whole sample of objects, computed considering the three 
physical implementations described in Section~\ref{ssec:feedback}.  When radiative 
cooling is included (\csfw\ and \csfwbh\ models) the level of clumpiness is very high and 
ranges from $\sim 3$ close to the centre up to $\sim10$ close to \rtwoh. When 
extending to outer regions the value of \crho\ increases exponentially reaching values 
of the order of 100 at $2 R_{200}$, with no significant difference when including BH 
feedback. When neglecting gas cooling (\ovisc), the values drop down by a factor of 5--10 
over the whole range, indicating that the value of \crho\ is mainly associated to small 
cold clumps.

However, when comparing our results with observational estimates one must consider 
that the cold dense gas at $T<10^5$ K does not produce any observable clumpiness in X-ray 
images. This makes the plotted \ovisc\ profile model a reference for the observed value 
of \crho\ without considering any detection of emitting clumps. For this reason we 
conclude that the value of $\sim$16 obtained by \cite{simionescu11} for the Perseus 
cluster is not realistic, since it is sufficient to exclude cold gas from our simulations 
to obtain significantly lower values.

The lower panel of Fig.~\ref{fig:clp_phys} shows the values of \crhov, which corresponds 
to the value of the clumpiness computed after applying the 1 per cent volume-clipping 
described in Section~\ref{ssec:volume}. With this method the difference associated to the 
physical implementations disappears almost completely, even for the \ovisc\ model: this 
indicates that 
large-scale inhomogeneities do not depend on the physical process that occur in the ICM, 
but on the intrinsic dynamical properties of the halo itself. In this case the median of 
\crhov\ computed on the whole sample has a value very close to 1 (e.g. almost uniform 
medium) close to the centre and grows constantly up to 1.3--1.4 at \rtwoh. There is a 
significant variance between the different haloes with about 25 per cent of objects 
having values of $\sim 1.4$. Outside \rtwoh\ the value of \crhov\ reaches $\sim 2$ at 
2\rtwoh, together with an increase of the dispersion between the objects: this 
reflects the intrinsic difference between the environment of the clusters' and groups' 
outskirts that can contain or not infalling haloes and accreting filaments.

This independence with respect to the physical implementation is partially in 
contrast with what obtained by \cite{zhuravleva13} that observe a slightly higher 
degree of inhomogeneity of the bulk in their NR runs, of the order of $\Delta C=0.1-0.2$. 
In fact, also in our simulations we see a small ($\Delta C \sim 0.04$) systematic excess 
in our \ovisc\ model, associated to the hot-dense tail of the ICM distribution. This 
discrepancy probably indicates a different cooling efficiency in the simulations analysed 
here and by \cite{zhuravleva13}.

\begin{figure}
\includegraphics[width=0.49\textwidth]{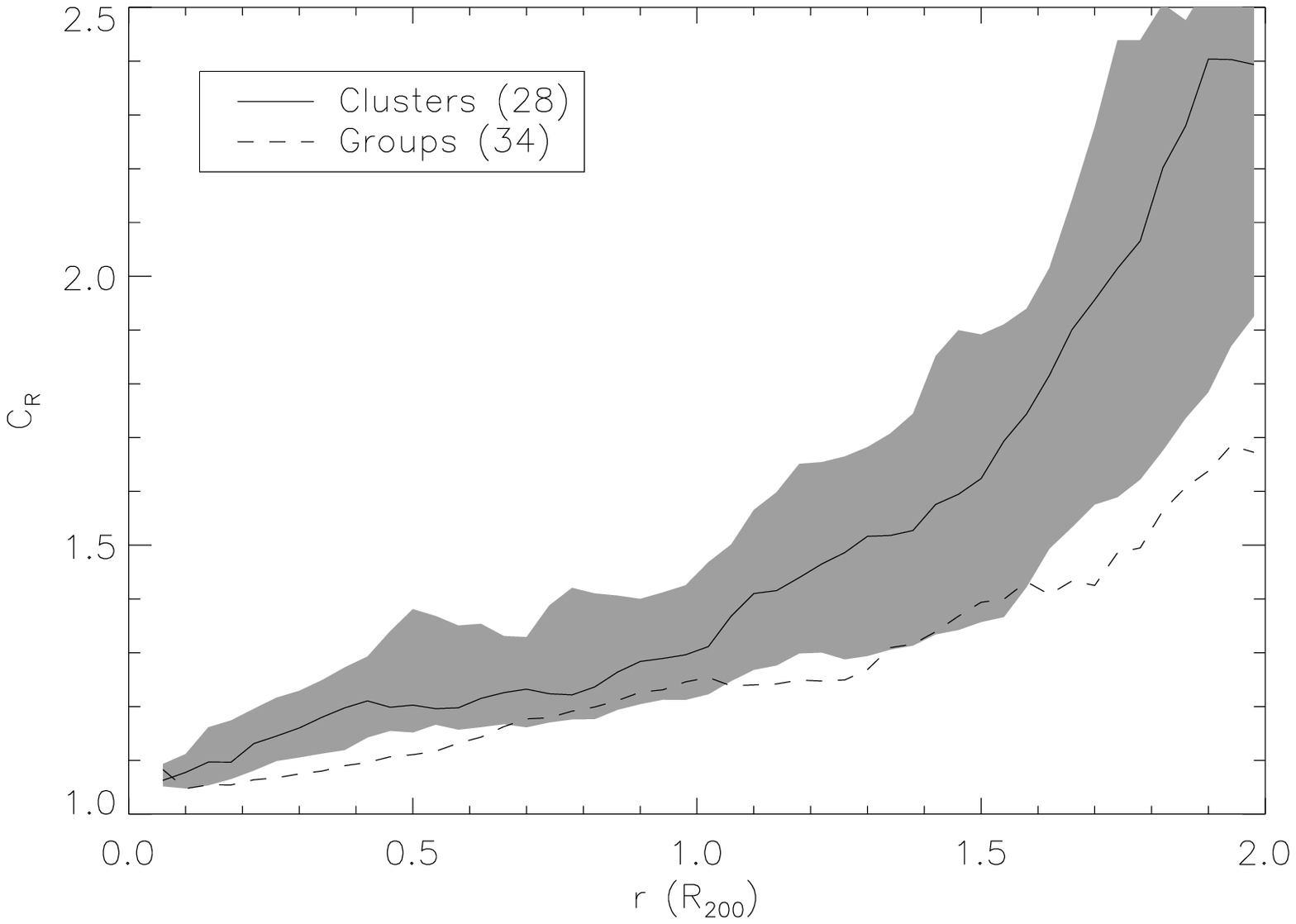}
\includegraphics[width=0.49\textwidth]{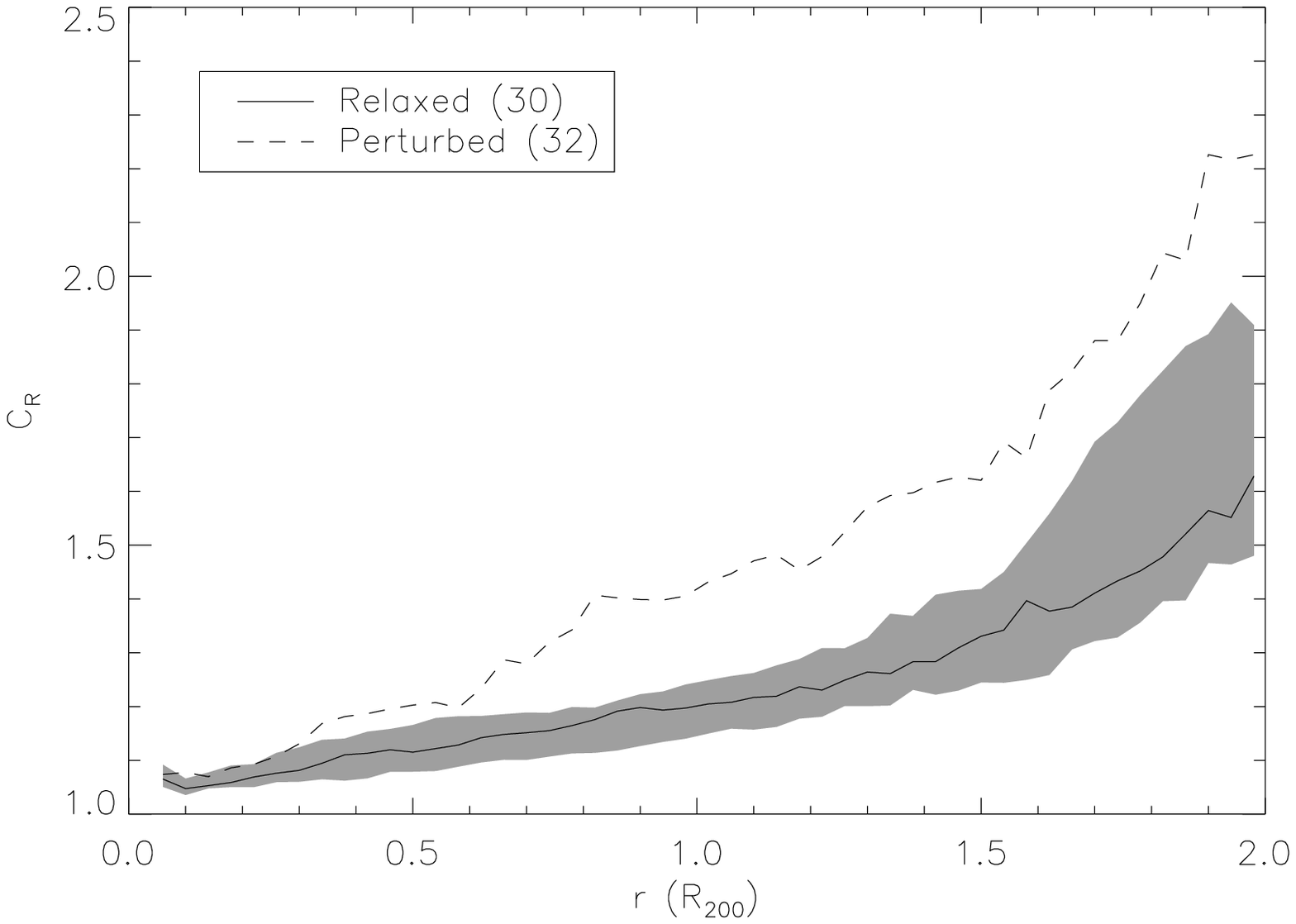}
\caption{Upper panel: residual gas clumpiness, i.e. after applying the volume-clipping 
method, for clusters (solid line) and groups (dashed line) as a function of distance from 
the centre for our reference model. The values represent the median of the sample (28 and 
34 objects, respectively), and the grey-shaded region encloses the quartiles of the 
clusters sample. Lower panel: same as upper panel but with relaxed (30 objects, solid 
line) and perturbed (32 objects, dashed line) haloes.}
\label{fig:clp_samp}
\end{figure}

It is interesting to see how the trend of \crhov\ varies when considering the mass of the 
halo and its dynamic status. We show in the top panel of Fig.~\ref{fig:clp_samp} the 
results of \crhov\ computed when separating the haloes according to their mass into 
groups and clusters (see Section~\ref{ssec:sample}). Clusters have on average a higher 
value of \crhov\, of the order of $\sim0.1$, with larger differences outside \rtwoh: this 
is due to the fact that massive objects have higher probability of accreting material 
even at later epochs, while galaxy groups are dynamically older, more dynamically 
stable and, therefore, more uniform. When separating relaxed and perturbed objects 
according to our definition (see Section~\ref{ssec:volume}) the difference is even more 
remarkable, as it can be seen in the bottom panel of Fig.~\ref{fig:clp_samp}. While close 
to the centre the two samples have very similar values of \crhov, the presence of 
accreting structures pushes the residual clumpiness of perturbed objects beyond 1.5 at 
\rtwoh. On the other side, relaxed haloes show all the same trend with \crhov\ raising 
linearly to 1.3--1.5 at \rtwoh: the residual clumpiness of these objects should be 
considered as an indicator of the amount of inhomogeneities common to all haloes, even 
the more dynamically stable, and connected to their departure from spherical symmetry 
and to the density fluctuations associated to the large-scale accretion patterns.

We observe as well that the differences in \crhov\ associated to the clusters/groups 
classification can be explained completely by the different correlation with the 
relaxed/perturbed one, (i.~e. clusters have higher values of \crhov\ because they are on 
average more perturbed than groups), making the latter classification the more 
interesting for our purposes. We note also that the trends that we observe are 
similar to those found by \cite{zhuravleva13}, (we refer to their 'bulk CSF' results), 
both in terms of radial dependence and with respect to the dynamical state of the 
haloes.


\section{The azimuthal scatter as a diagnostic of the residual clumpiness}
\label{sec:scatter}

\subsection{The azimuthal scatter}
As said, the clumpiness of the ICM is not a directly measurable quantity. Here we 
investigate the possibility of obtaining an estimate from the observed azimuthal scatter 
of the X-ray and SZ profiles of the haloes.

Following \cite{vazza11}, we define the azimuthal scatter of an observable quantity $x$ 
as
\begin{equation}
\sigma_x (r) \equiv \sqrt{\frac{1}{N} 
             \sum_{i=1}^N \left( \frac{x_i(r)-\bar{x}(r)}{\bar{x}(r)} \right)^2 } \ ,
\label{eq:scatter}
\end{equation}
where $N$ is the number of azimuthal sectors, $x_i(r)$ is the radial profile of the 
quantity in a given sector and $\bar{x}(r)$ is the average over the $N$ sectors at 
distance $r$ from the centre. In our case we fix $N=12$, so that each sector 
corresponds to 30 degrees: this value is enough to account for all of the main azimuthal 
fluctuations associated to typical ICM inhomogeneities \citep[see the discussion 
in][]{vazza11}. For the purpose of our work we consider the scatter in the X-ray surface 
brightness (computed in different bands) and in the thermal SZ effect.

We compute the X-ray surface brightness for every halo in the same 50 radial bins and 
in the 12 different sectors by assuming an \textsc{apec} emission model \citep{smith01}, 
by fixing the redshift\footnote{Since we are interested in the azimuthal scatter of the 
X-ray surface brightness and not in its absolute value, fixing the redshift has effect 
only in the definition of the X-ray bands.} of our clusters to $z=0$. Since our \csfwbh\ 
model follows also the enrichment of metals with the recipe of \cite{tornatore07}, we 
consider also the contribution from different chemical species in the computation of the 
emissivity of the SPH particles \citep[the details of the procedure are described 
in][]{roncarelli12}. For what concerns the SZ scatter, the method to compute the value of 
the \ypar\ from our SPH simulation is the same as in \cite{roncarelli07}. For these 
computations we are considering all the SPH particles, without applying the 
volume-clipping method (see Section~\ref{ssec:volume}).

\begin{figure}
\includegraphics[width=0.49\textwidth]{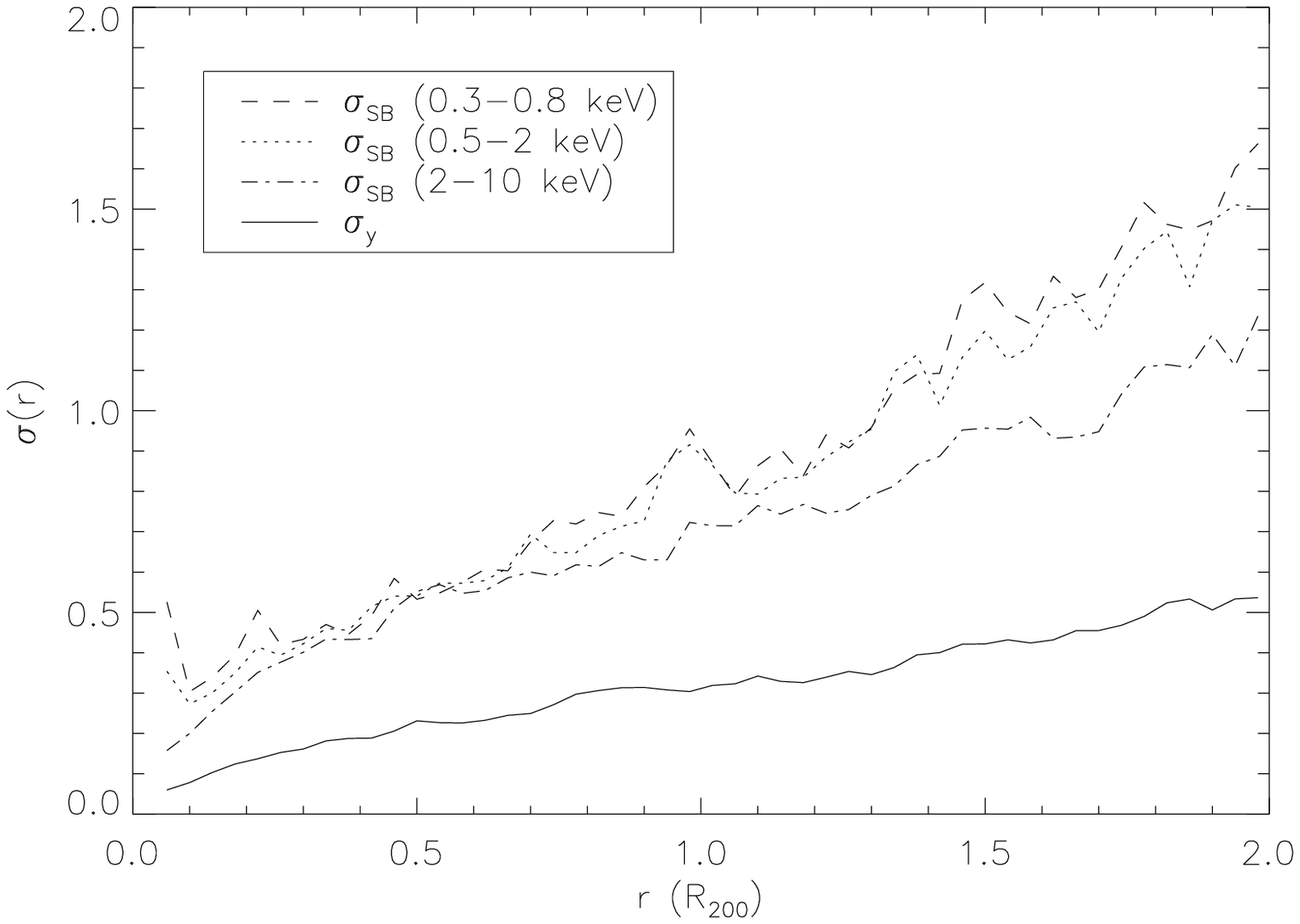}
\includegraphics[width=0.49\textwidth]{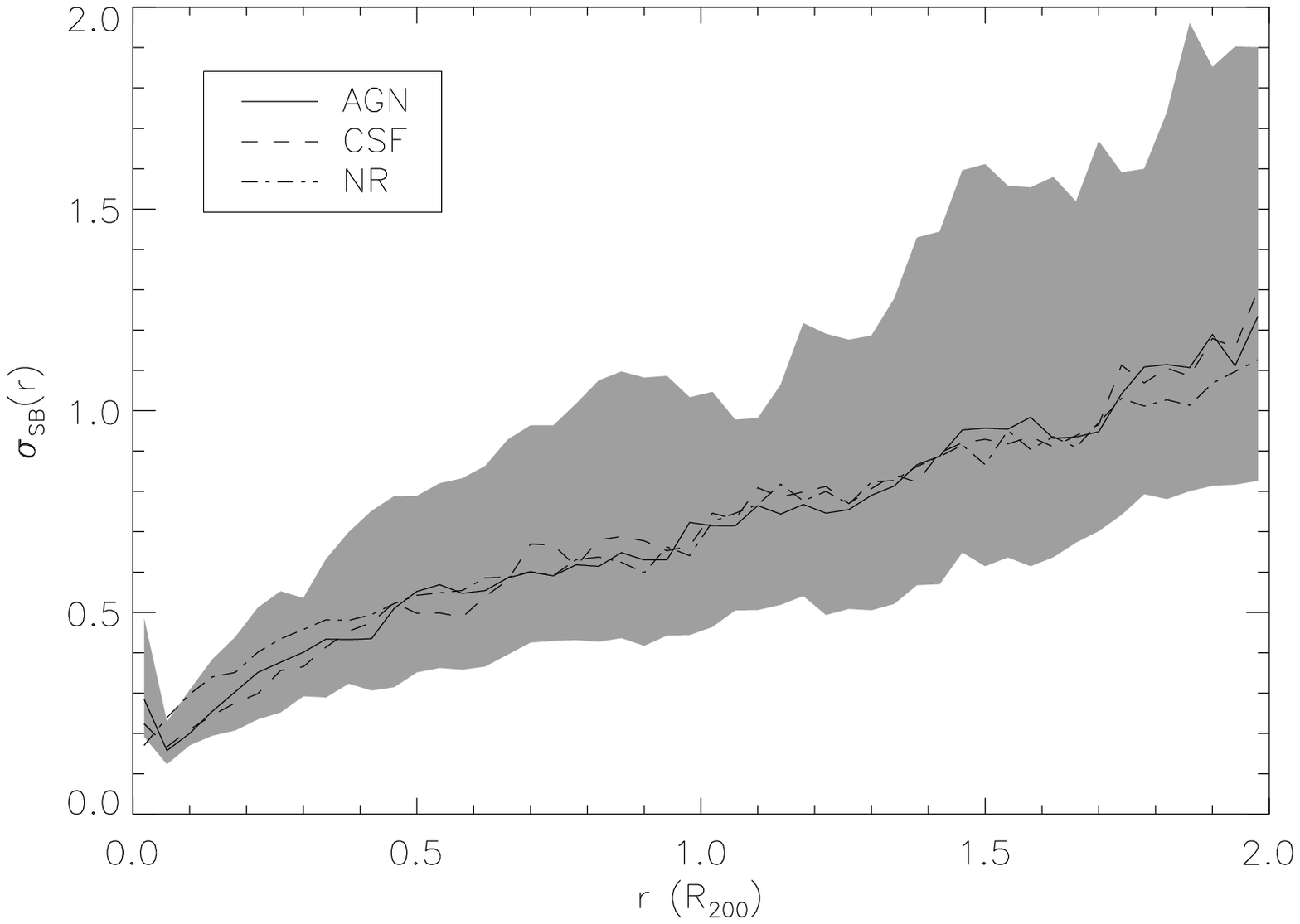}
\caption{Upper panel: azimuthal scatter of the \ypar\ (solid line) and of the X-ray 
surface brightness in the 0.3--0.8, 0.5--2 and 2--10 keV bands (dashed, dotted and 
dot-dashed line, respectively) as a function of distance from the cluster centre. The 
values represent the median computed over the whole sample of 62 haloes in 12 azimuthal 
sectors. Lower panel: same as upper panel but for the 2--10 keV surface brightness only, 
computed with three different physical implementations: \csfwbh\ (solid line), \csfw\ 
(dashed) and \ovisc\ (dot-dashed). The grey-shaded area encloses the quartiles of the 
clusters sample for our reference model.}
\label{fig:sc_phys}
\end{figure}

\begin{figure}
\includegraphics[width=0.49\textwidth]{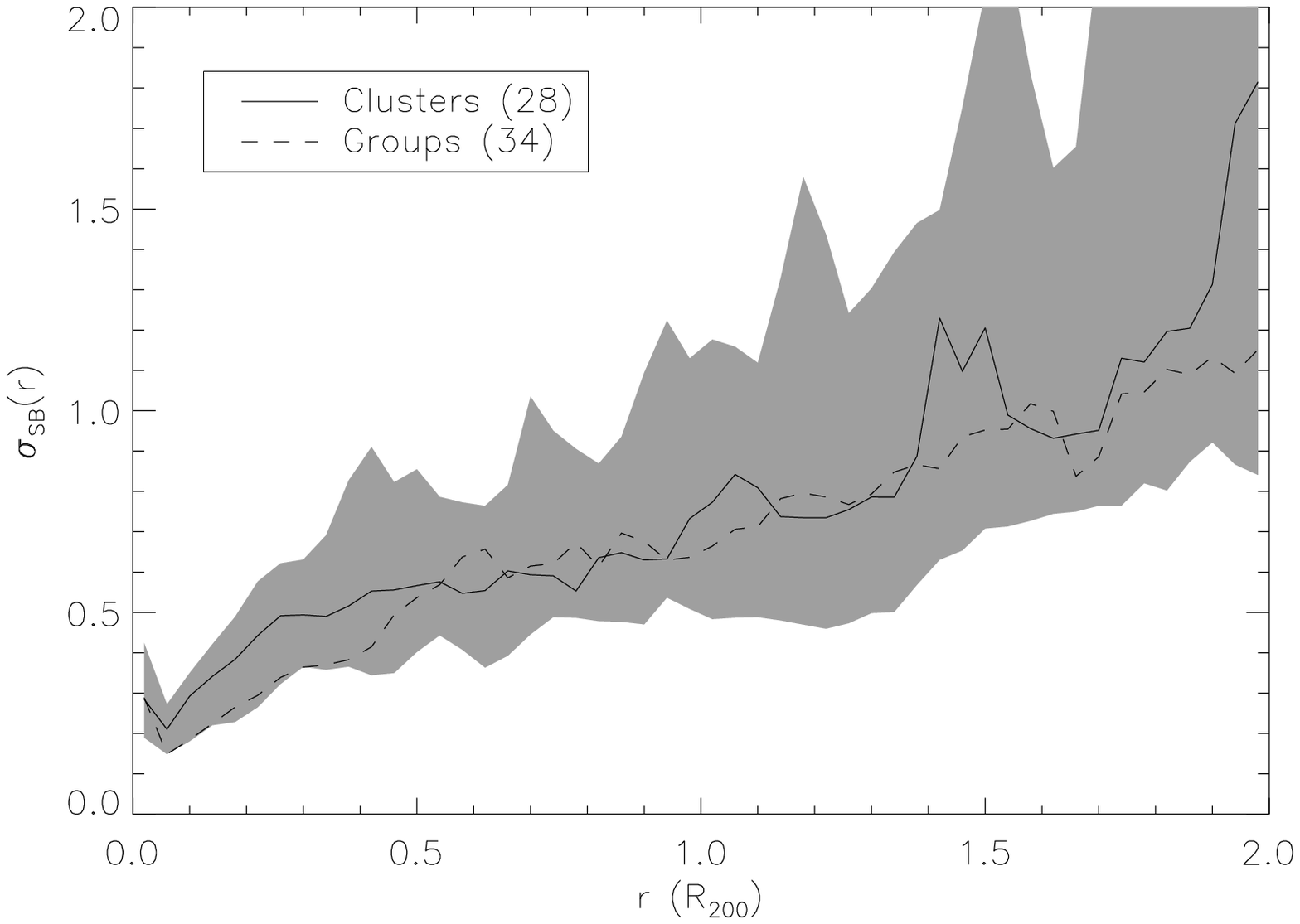}
\includegraphics[width=0.49\textwidth]{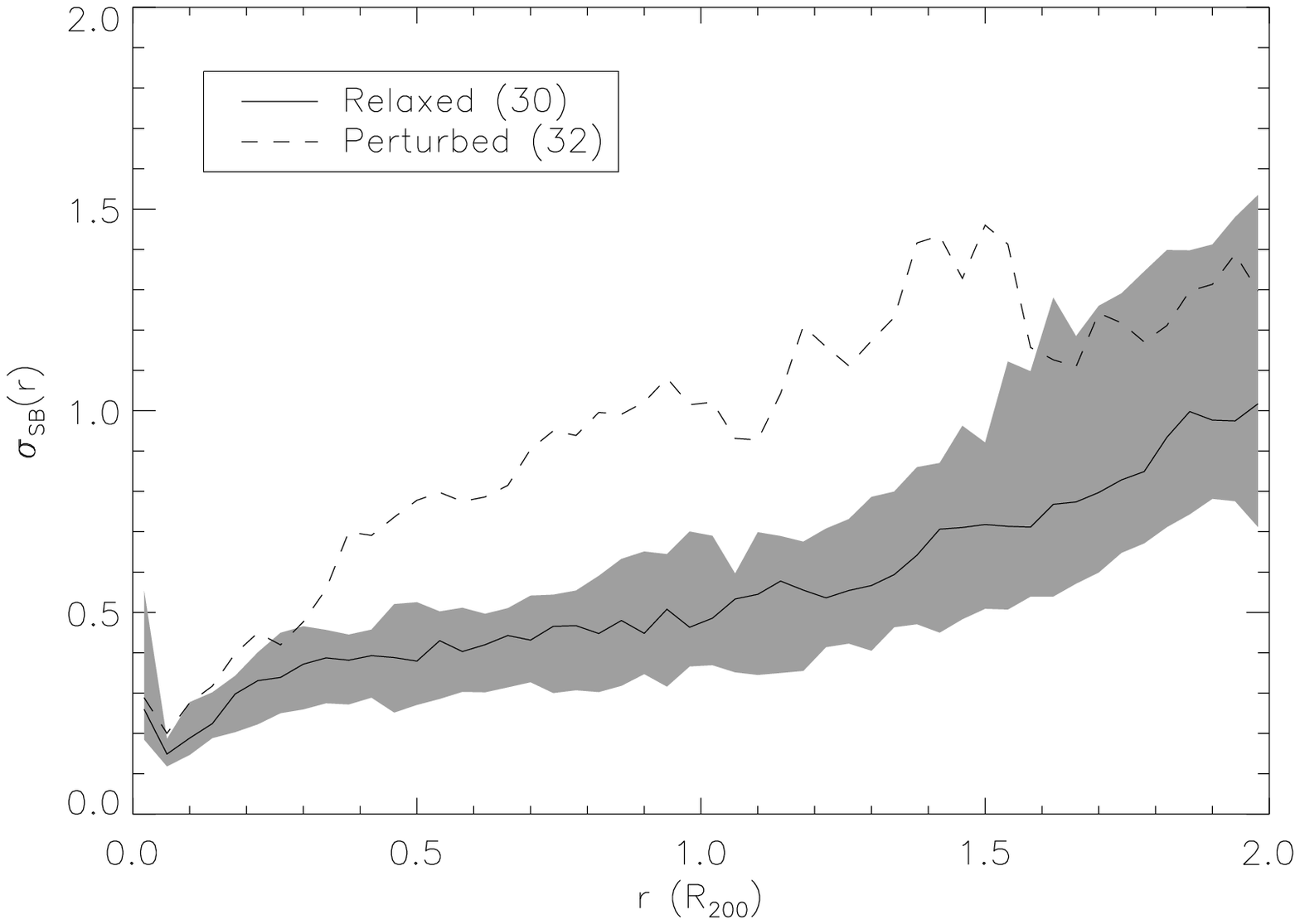}
\caption{Upper panel: azimuthal scatter of the 2--10 keV surface brightness for clusters 
(solid line) and groups (dashed) as a function of projected distance from the centre for 
our reference model. The values represent the median of the two samples (28 and 34 
objects, respectively), and the grey-shaded region encloses the quartiles of the clusters 
sample. Lower panel: same as upper panel but with relaxed (30 objects, solid line) and 
perturbed (32 objects, dashed line) haloes.}
\label{fig:sc_samp}
\end{figure}

We show in the upper panel of Fig.~\ref{fig:sc_phys} the azimuthal scatter as a function 
of distance from the 
centre for the value of the \ypar\ and the surface brightness in three different X-ray 
bands. At each distance we show the median computed over the whole sample of 62 haloes 
in the \csfwbh\ model. The general observed trend is an increase of the azimuthal scatter 
with radius, with the value of $\sigma_y$ going from $\sim \,$0 at the centre to 
$\sim \,$0.3 at \rtwoh. The scatter associated to the X-ray surface brightness is 2--3 
times higher at all distances, with lower energy bands (0.3--0.8 and 0.5--2 keV) having 
higher values with respect to the hard 2--10 keV band, indicating that these 
inhomogeneities are associated to gas at temperatures around 0.5--1 keV, that has the 
peak of its emission at soft X-ray energies.

It is interesting to note that when studying how the azimuthal scatter varies with 
respect to the assumed physics and to the halo classifications, the trend is very similar 
with respect to \crhov, rather than the global clumping factor \crho\ (refer to 
Section~\ref{ssec:crho_samp}). For instance, the lower panel of Fig.~\ref{fig:sc_phys} 
shows that the value of \sigmasb\ is almost independent of the assumed physical model. 
When looking at the dependence on the mass and the dynamical status of the halo (upper 
and lower panels of Fig.~\ref{fig:sc_samp}, respectively) we can see that on average 
galaxy clusters show a slightly higher azimuthal scatter, while perturbed systems 
have values of $\sigma$ which are higher by a factor of 2 with respect to relaxed ones at 
\rtwoh.


\subsection{Correlating the clumpiness with the azimuthal scatter}
Since we have shown that clumpiness and azimuthal scatter have similar radial trends, 
here we address directly the question whether a tight correlation exists between these 
variables and if there exists a possible relation that can connect one with the other.

To this purpose we perform a statistical analysis of these variables by computing the 
Spearman's rank correlation, $r_{\rm S}$, of \crho\ and of \crhov\ with the scatter in 
the X-ray and SZ profiles of our clusters. We restrict our analysis to the interval 
$0.15<r/$\rtwoh$<1.5$, in order to cut out the core, and to the sample of the relaxed 
haloes for which the definition of \crhov\ is more robust. Given the large sample of 
haloes available, these restrictions do not affect the robustness of our computation. 
The results are shown in Table~\ref{tab:r_s}.

\begin{table}
\begin{center}
\caption{Spearman's rank correlation ($r_{\rm S}$) of the values of \crho\ and \crhov\ 
(second and third column, respectively) with the azimuthal scatter in the X-ray surface 
brightness in different bands (second to fourth row, respectively) and in the \ypar\ 
profiles (fifth row). The values analysed are taken from the 30 relaxed haloes 
considering the bins in the range $0.15<r/$\rtwoh$<1.5$.}
\begin{tabular}{lccc}
\hline
\hline
                                 & & \crho  & \crhov \\
\hline
$\sigma_{\rm SB}$ (0.3--0.8 keV) & & 0.32   & 0.59   \\
$\sigma_{\rm SB}$ (0.5--2 keV)   & & 0.30   & 0.60   \\
$\sigma_{\rm SB}$ (2--10 keV)    & & 0.23   & 0.61   \\
$\sigma_y$                       & & 0.35   & 0.68   \\
\hline
\hline
\label{tab:r_s}
\end{tabular}
\end{center}
\end{table}

The main conclusion that can be drawn from these values is that the azimuthal scatter of 
both X-ray and SZ profiles has a very high degree of correlation with the values of 
\crhov, i.~e. with the inhomogeneities on the large scales: in fact, the values of 
$r_{\rm S} = 0.6-0.7$ indicate that it is possible to describe \crhov\ as a monotonic 
increasing function of $\sigma_y$ and $\sigma_{\rm SB}$. On the other side when 
considering the correlation of the different azimuthal scatters with the total 
clumpiness, \crho, the correlation still exists although being much weaker. It is worth 
to point out also how the trend with the X-ray energy is opposite in the two cases. Since 
clumps embed lower temperature ICM, softer bands are more correlated to \crho\ with 
respect to the hard one. On the other side, when clumps are excised, the densest gas is 
also hotter thus the correlation \crhov--\sigmasb\ increases at higher energy.

\begin{figure}
\includegraphics[width=0.49\textwidth]{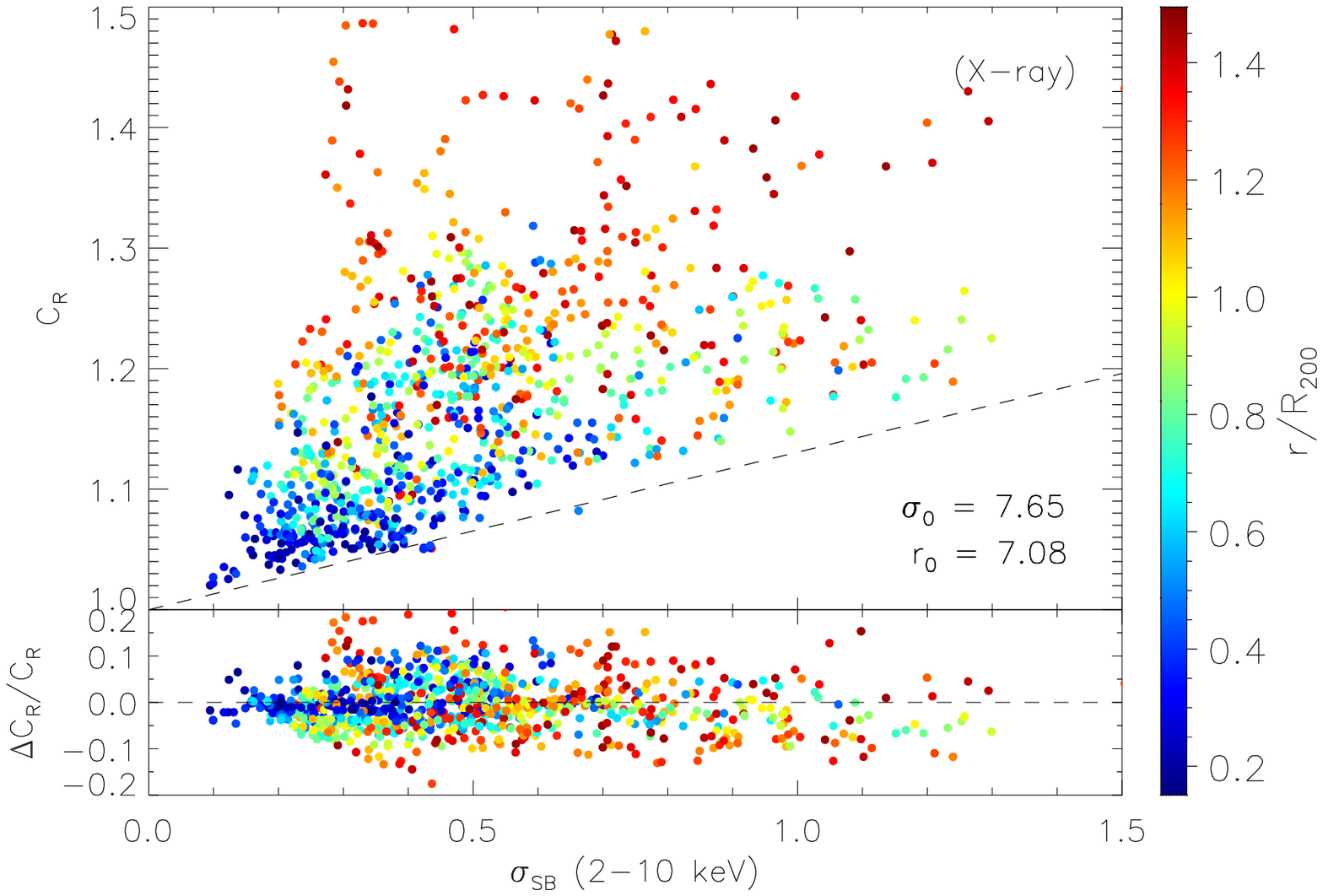}
\includegraphics[width=0.49\textwidth]{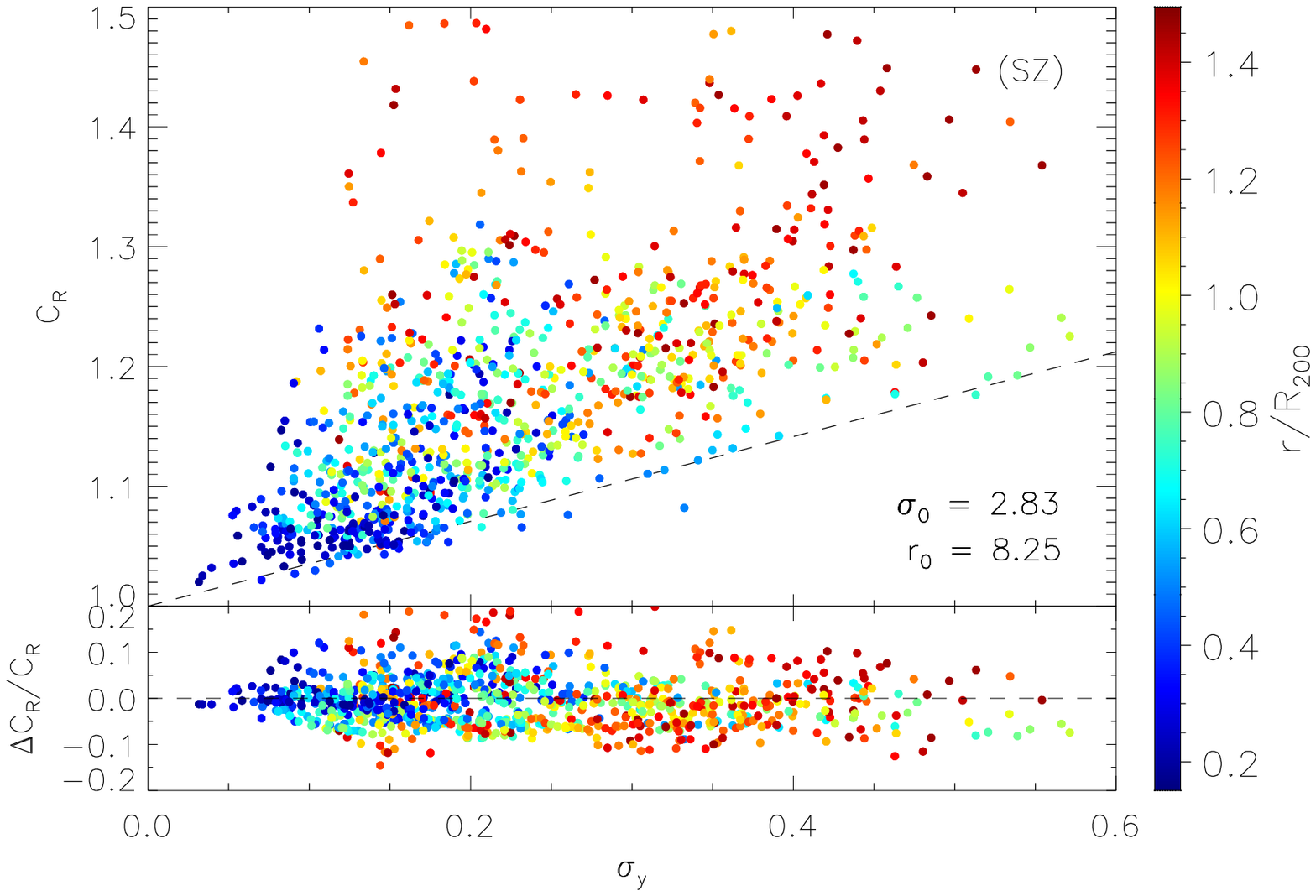}
\caption{Upper panel: correlation between the azimuthal scatter in the 2--10 keV band 
and \crhov. The points considered correspond to the sample of the 30 relaxed haloes 
in the range $0.15<r/$\rtwoh$<1.5$ (a total of 990 points, 33 for each halo) and their 
colors indicate the radial distance, increasing from blue to red. The quantities 
indicated ($\sigma_0$ and $r_0$  in units of \rtwoh) are the best-fit values obtained 
using the formula of eq.~(\ref{eq:fitfun}) and the lower part shows the residuals with 
respect to the model in terms of relative difference in the clumpiness value. The 
diagonal dashed 
line indicates the relation \crhov$=1+\sigma/\sigma_0$, i.e. the best-fit formula in the 
limit of $r=0$. Bottom panel: same as upper panel but for the scatter in the \ypar.}
\label{fig:fit}
\end{figure}

More in the detail, we show in Fig.~\ref{fig:fit} the correlation between \crhov\ 
and the scatter in the 2--10 keV band (top panel): although there is a significant 
dispersion, the trend of increasing clumpiness with increasing value of $\sigma_{\rm SB}$ 
is clear. The same applies also when considering the scatter in the \ypar\ (bottom 
panel). When analysing the dispersion of the relation, it is clear also the trend 
of higher values of \crhov\ for increasing radii (indicated by the different colors), as 
already discussed for Figs.~\ref{fig:clp_phys} and \ref{fig:clp_samp}.

Since it is not possible to define \emph{a priori} the relation between scatter and 
clumpiness, we proceed empirically by introducing the following function

\begin{equation}
C_\mathcal{R}^{\rm est} (\sigma,r)= 1 + \frac{\sigma}{\sigma_0} + \frac{r}{r_0} \,
\label{eq:fitfun}
\end{equation}
that connects the azimuthal scatter $\sigma$ and the distance from the center $r$ with an 
estimate of the clumpiness \crhoest. The choice of this expression 
is made in order to have a simple increasing function of both $\sigma$ and $r$ (which 
holds for $\sigma_0>0$ and $r_0>0$), and \crhoest$(\sigma=0,r=0)=1$. 

Given the two free parameters, $\sigma_0$ and $r_0$, for every observable quantity object 
of our analysis (i.e. azimuthal scatter in the X-ray surface brightness and in the \ypar) 
we determine their best-fit values by fitting\footnote{In the 
fitting procedure we omit to assign errors to our data.} the expression of 
eq.~(\ref{eq:fitfun}) using the points displayed in Fig.~\ref{fig:fit}, i.e. \crhov\ as a 
function of $\sigma$ and $r$. Again, we restrict this calculation to the 30 relaxed 
clusters and to the range $0.15<r/$\rtwoh$<1.5$. The results are shown in 
Fig.~\ref{fig:fit}, together with the residuals, for the 2--10 keV band and for the 
\ypar. It is clear that the best-fit function provides a good global description of the 
relation \crhov--$(\sigma,r)$, with almost all of the points that have an estimate within 
10 per cent of the true value. Furthermore, we note how the diagonal dashed line 
corresponding to best-fit relation in the limit of $r=0$ defines well the ``forbidden'' 
region of the \crhov--$\sigma$ plane, below the line itself. Table~\ref{tab:bestfit} 
shows the best-fit values for all the three X-ray bands and for the thermal SZ effect.

\section{Improving mass estimates}
\label{sec:corr_mass}

We have shown in the previous section that it is possible to obtain an estimate of the 
large-scale inhomogeneities (\crhov) from the azimuthal scatter in the X-ray surface 
brightness and \ypar. The question that arises is whether it is possible to use the 
observed value of $\sigma(r)$ as a function of the distance from the cluster centre, to 
improve the estimates of the gas density profile and, consequently, the measurement of 
\mgas\ and \mhe.


\subsection{Correcting the gas density profile}
\label{ssec:corr_rho}

The estimate of the gas density profile $\rho(r)$ from the X-ray surface brightness is 
affected by the problem that the corresponding emissivity $\epsilon_X(r)$ in a fixed band 
depends on the squared gas density: $\epsilon_X \propto \, <\rho^2>_V$. Given the 
definition of eq.~(\ref{eq:clp0}), an observer that ignores the residual clumpiness of 
the gas and assumes spherical symmetry obtains

\begin{equation}
\rho_{\rm X}(r) = \sqrt{C_\mathcal{R}(r)} \, \rho(r) \ , 
\label{eq:rhox}
\end{equation}
which corresponds to a systematic overestimate.

If we consider the observed azimuthal scatter profile, $\sigma(r)$, and use the relation 
of eq.~(\ref{eq:fitfun}) to obtain an estimate of \crhov, we can correct our measurement 
of the density profile as follows
\begin{equation}
\hat\rho(r) = \frac{\rho_{\rm X}(r)}{\sqrt{C_\mathcal{R}^{\rm est}(\sigma,r)}}  \ . 
\label{eq:crho}
\end{equation}

\begin{figure*}
\includegraphics[width=0.49\textwidth]{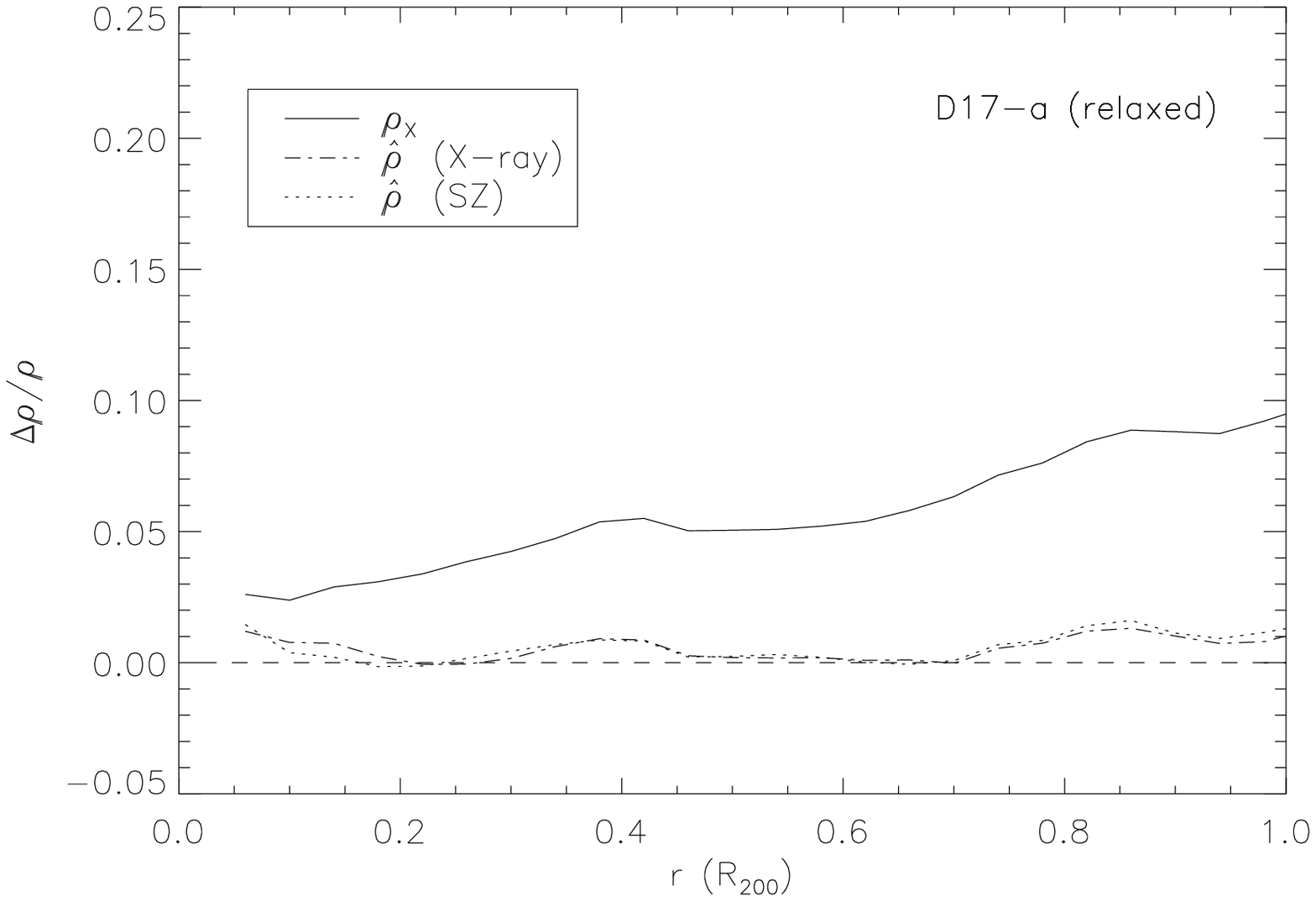}
\includegraphics[width=0.49\textwidth]{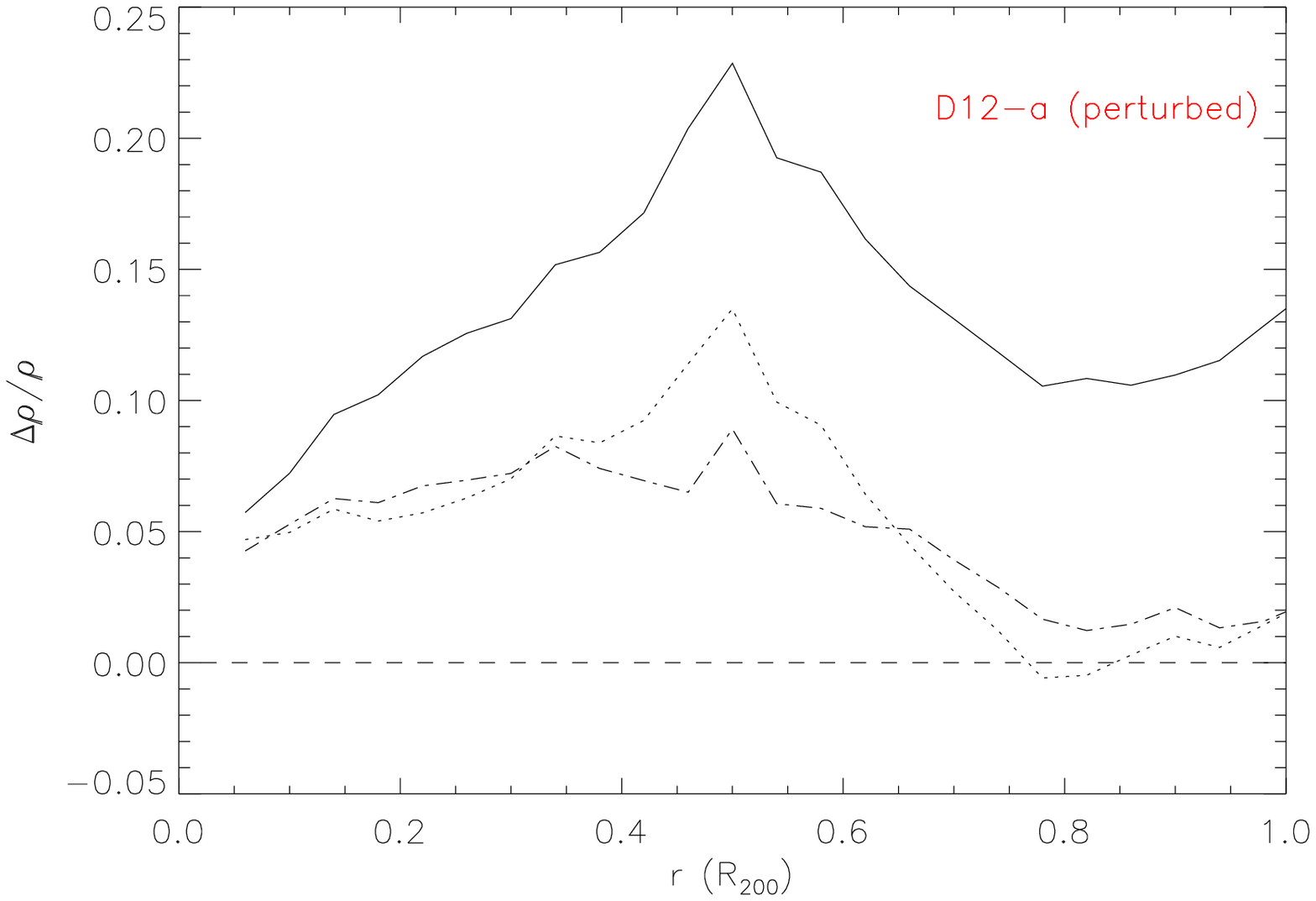}
\caption{Sketch of the gas density correction method applied to the two clusters shown in 
the maps of Fig.~\ref{fig:maps}: D17-\emph{a} (left panel) and D12-\emph{a} (right 
panel). We show the radial dependence of the relative differences between the true gas 
density and the uncorrected measured density (solid line), the density after applying the 
X-ray and SZ corrections (dot-dashed and dotted line, respectively). The horizontal 
dashed line shows the ideal case of perfect density profile reconstruction 
($\Delta\rho=0$).}
\label{fig:rho_corr}
\end{figure*}

As examples, we show in Fig.~\ref{fig:rho_corr} the application of this method to the two 
clusters of Fig.~\ref{fig:maps}. The relaxed D17-\emph{a} cluster (left panel) shows 
reconstructed density profiles very close to the true value\footnote{Here we are 
considering the density  profiles after applying the volume-clipping method described in 
Section~\ref{ssec:volume}, so we are implicitly assuming that small clumps have been 
efficiently removed in X-ray observations.}, with differences smaller than 
2 per cent up to \rtwoh. On the other side, the corresponding uncorrected X-ray profile 
overestimates the true value by 5--10 per cent over the whole $r > 0.4$\rtwoh\ range.

When looking at the D12-\emph{a} cluster (right panel), the presence of the disturbing 
structure still produces an overestimate of the corrected density profiles of about 5--10 
per cent in correspondence of the clumpiness peak. However, even in this case the 
improvement with respect to the original $\rho_{\rm X}$ profile overestimate is 
remarkable, through all the radial range.

This method can be directly applied to the observed density profiles and scatter of 
\cite{eckert12}. In fact, they obtained a measurement of \sigmasb\ in the soft 0.5--2 keV 
band in their 31 ROSAT-PSPC objects (we refer to their Fig.~7, entire sample) up to 
\rtwoh. By using this quantity to estimate the residual clumpiness, we obtain an 
approximate value of \crhoest$=1.2$ at $r\sim$\rtwoh, that corresponds to an overestimate 
of $\sim8$ per cent in the gas density. This consideration leads to mitigate the existing 
conflict between their observed density profiles and the simulated ones, although not 
enough to solve the problem completely.


\subsection{Correcting the gas mass bias}
\label{ssec:corr_mgas}
If its gas density profile $\rho(r)$ is known, then the gas mass of the halo enclosed by 
\rtwoh\ can be obtained from the following formula:
\begin{equation}
M_{\rm gas} = 4\pi \int_0^{R_{200}} \rho(r) \, r^2 dr \ .
\label{eq:mgas}
\end{equation}

However, the density bias described by eq.~(\ref{eq:rhox}) reflects also into the gas 
mass estimate,

\begin{equation}
M_{\rm gas,X} = 4\pi \int_0^{R_{200}} \sqrt{C_\mathcal{R}(r)} \rho(r) \, r^2 dr \ ,
\label{eq:mbias}
\end{equation}
that leads to a bias in the gas mass measurement

\begin{equation}
b(M_{\rm gas}) \equiv \frac{M_{\rm gas,X}-M_{\rm gas}}{M_{\rm gas}}
\label{eq:cbias}
\end{equation}
which is always $> 0$ reaching null value only in the limit of a perfectly uniform medium 
(\crhov$=1$).

\begin{figure}
\includegraphics[width=0.49\textwidth]{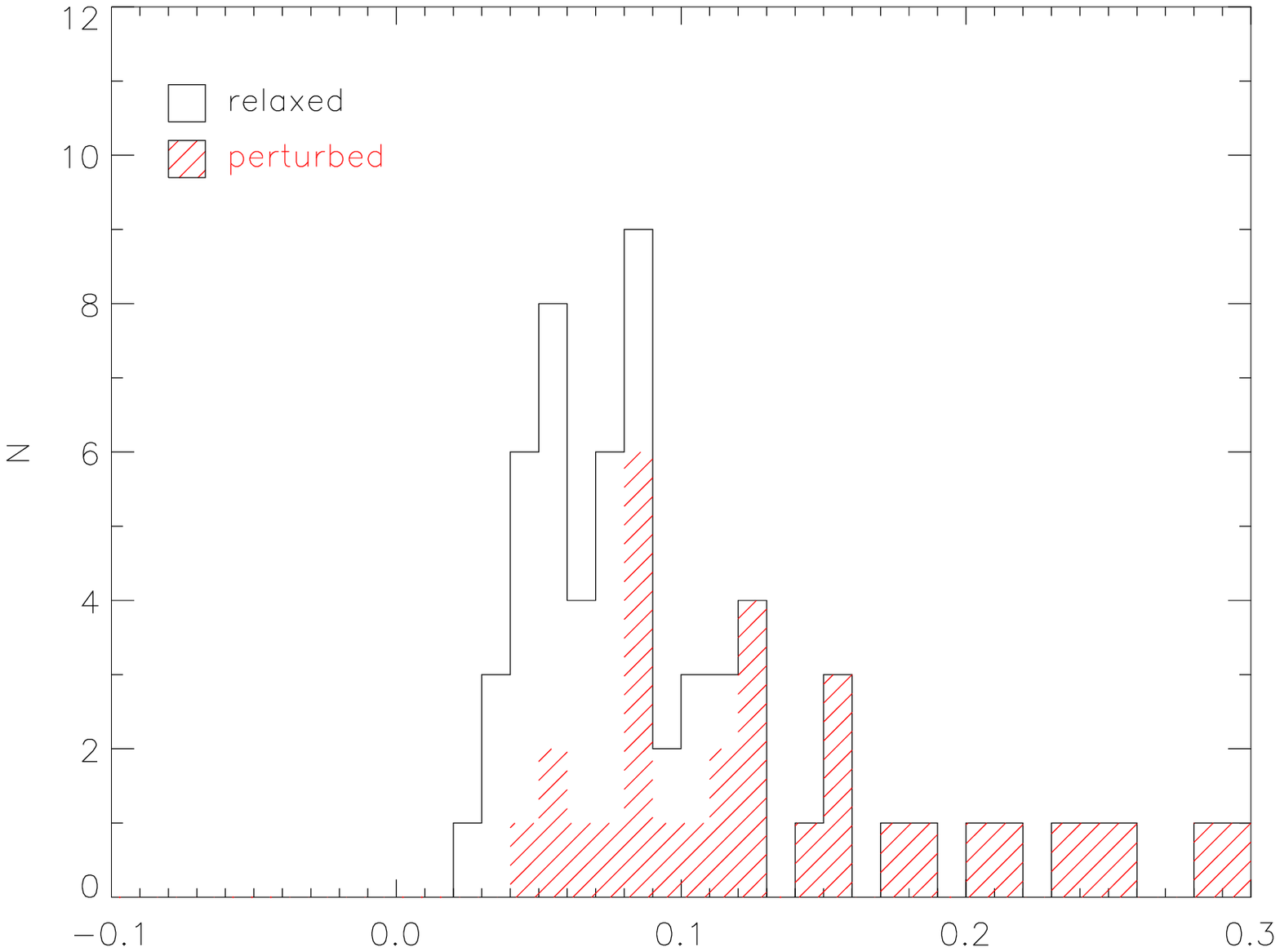}
\includegraphics[width=0.49\textwidth]{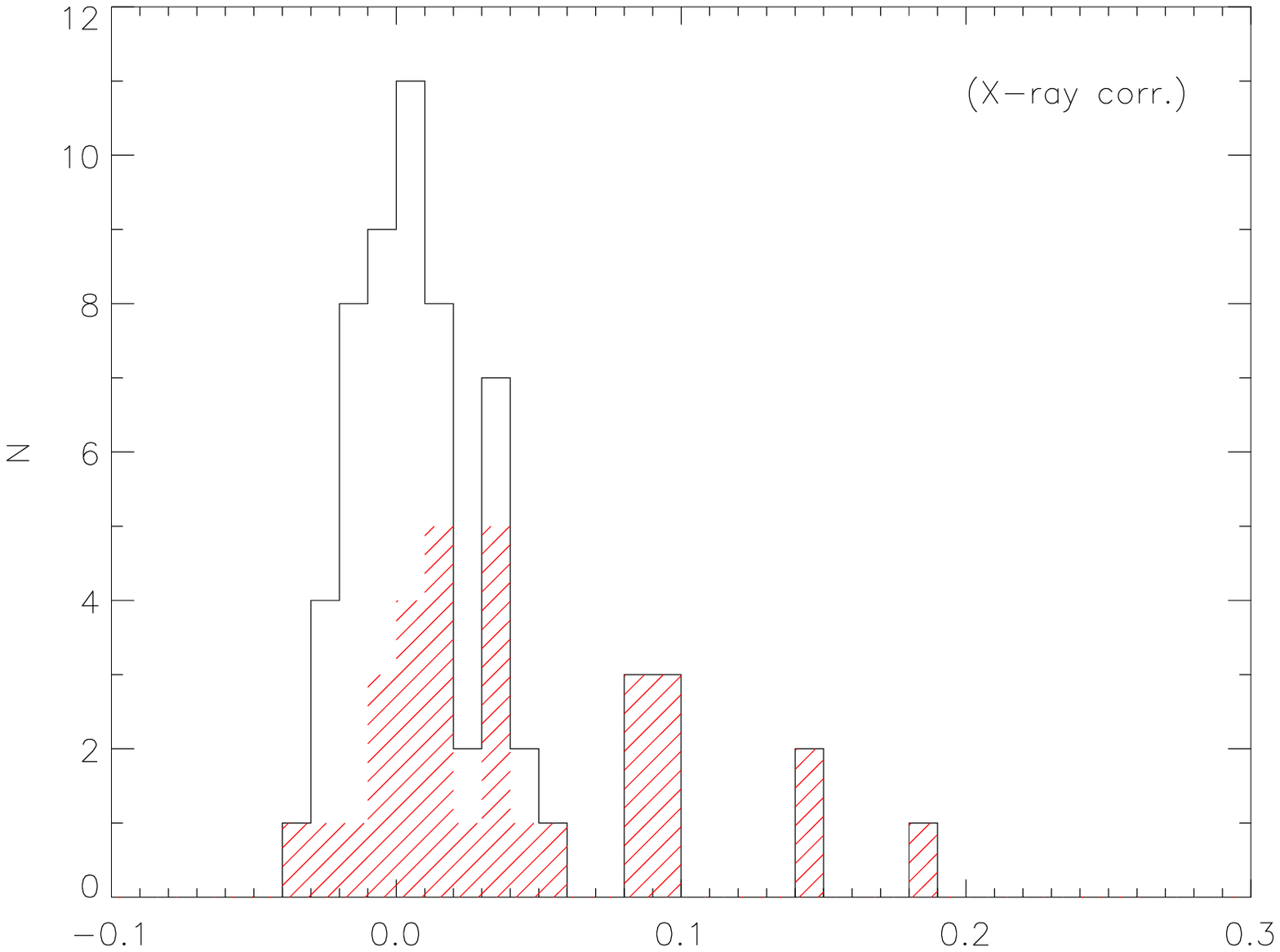}
\includegraphics[width=0.49\textwidth]{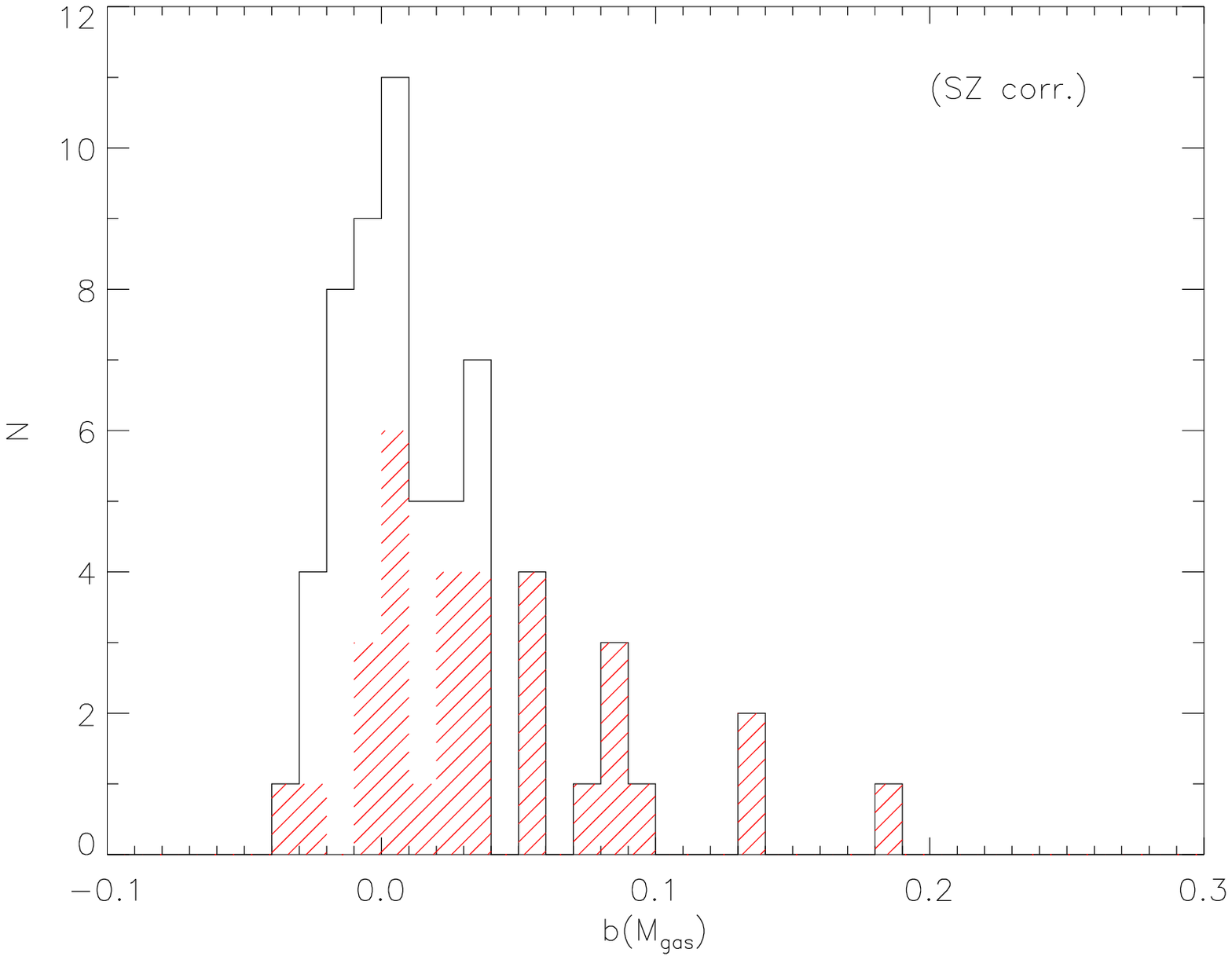}
\caption{Top panel: histogram of the estimated gas mass bias associated to the residual 
clumpiness for our sample of 62 clusters. Perturbed haloes are marked by the shaded 
areas. Central and bottom panels: same as top panel but after the 
correction obtained by estimating the value of \crhov\ from the scatter in the 2--10 keV 
surface brightness and \ypar, respectively.}
\label{fig:bias}
\end{figure}

Given our set of simulated clusters and groups, we can obtain the expected value of 
$b(M_{\rm gas})$ from their density and clumpiness profiles. The results are 
shown in the top panel of Fig.~\ref{fig:bias}: typical values correspond to 
overestimates of 5--10 per cent, with the more perturbed haloes that can reach 
overestimates around 20--30 per cent.

If we substitute in eq.~(\ref{eq:mgas}) the corrected density profile of 
eq.~(\ref{eq:crho}), we obtain a new estimate of the gas mass
\begin{align}
\widehat{M}_{\rm gas} & = 4\pi \int_0^{R_{200}} \hat{\rho}(r) \, r^2 dr = \nonumber \\
                      & = 4\pi \int_0^{R_{200}}
                    \sqrt{\frac{C_\mathcal{R}(r)}{C_\mathcal{R}^{\rm est}(\sigma,r)}} 
                    \, \rho(r) \, r^2 dr
\label{eq:corr_mass}
\end{align}
which provides a good approximation of \mgas\ as much as \crhoest\ is an accurate 
estimate of the residual clumpiness.

We verify the accuracy of this method by applying it to our simulated haloes. The central 
and bottom panels of Fig.~\ref{fig:bias} show the distribution of the 
bias after applying the correction considering the scatter in the 2--10 keV band surface 
brightness and in the \ypar, respectively. We report in Table~\ref{tab:bestfit} the 
values of the median of the corrected and uncorrected bias distributions together with 
their dispersion also for the other two X-ray bands, for our sample of 30 relaxed haloes. 
In all cases the systematics for relaxed systems is completely eliminated, with the halo 
to halo scatter which is reduced as well. A very small number of highly perturbed haloes 
keep having overestimates of 10--20 per cent: this reflects the presence of other types 
of asymmetries in the profiles not directly associated to the clumpiness of the ICM. 
Nevertheless, when calculating the median and the quartiles of the whole sample 
distribution, as shown in Table~\ref{tab:bestfit_all}, we can see that our method is 
still effective even if applied ``blindly'' to all objects.

\begin{table*}
\begin{center}
\caption{Best-fit values of the parameters $\sigma_0$ and $r_0$ (second and third 
column, respectively, with $r_0$ in units of \rtwoh) of the relation 
$C_\rho^{\rm est}(\sigma,r)$, being $\sigma$ the 
azimuthal scatter of the surface brightness in the three X-ray bands (second to fourth 
row) and the scatter of the \ypar\ profiles (fifth row). The fourth column shows the 
median (in percent units) of the distributions of bias in the value of \mgas\ at \rtwoh\ 
for our sample of 30 relaxed haloes, together with the difference with respect to the 
upper and lower quartiles (indicated as error), after the clumpiness correction of 
eq.~(\ref{eq:corr_mass}). The fifth and sixth column show the same quantities for the 
distribution of \mhe\ and \fgas, respectively. In the last row we report the values 
corresponding to the uncorrected biases.}
\begin{tabular}{lcccccc}
\hline
\hline
                                 & & $\sigma_0$ & $r_0$ & $b(M_{\rm gas})$ \% &  $b(M_{\rm he})$ \%&  $b(f_{\rm gas})$ \% \\
\hline
\vspace{0.1cm}
$\sigma_{\rm SB}$ (0.3--0.8 keV) & &   22.85     &    5.51    & $+0.02_{- 0.85} ^{+ 1.27}$ & $-0.56_{- 1.19} ^{+ 1.68}$ & $+0.26_{- 1.31} ^{+ 2.34}$ \\
\vspace{0.1cm}
$\sigma_{\rm SB}$ (0.5--2 keV)   & &   16.02     &    5.87    & $+0.08_{- 0.94} ^{+ 1.22}$ & $-0.38_{- 1.19} ^{+ 1.69}$ & $+0.06_{- 0.98} ^{+ 2.38}$ \\
\vspace{0.1cm}
$\sigma_{\rm SB}$ (2--10 keV)    & &    7.65     &    7.08    & $-0.58_{- 0.67} ^{+ 1.53}$ & $+0.55_{- 2.63} ^{+ 1.87}$ & $-0.45_{- 1.34} ^{+ 2.59}$ \\
$\sigma_y$                       & &    2.83     &    8.25    & $-0.34_{- 0.89} ^{+ 1.40}$ & $+0.95_{- 1.47} ^{+ 1.84}$ & $-1.59_{- 1.27} ^{+ 3.02}$ \\
\hline
No correction                    & &             &            & $+6.11_{- 1.51} ^{+ 1.73}$ & $-2.22_{- 2.02} ^{+ 1.52}$ & $+8.45_{- 2.11} ^{+ 2.57}$ \\
\hline
\hline
\label{tab:bestfit}
\end{tabular}
\end{center}
\end{table*}

\begin{table*}
\begin{center}
\caption{Same as Table~\ref{tab:bestfit} but for the whole sample of 62 haloes. The best-fit $\sigma_0$ and $r_0$ are not quoted here since we 
assume the same values of the relaxed sample.}
\begin{tabular}{lcccc}
\hline
\hline
                                 & &  $b(M_{\rm gas})$ \% &  $b(M_{\rm he})$ \%&  $b(f_{\rm gas})$ \% \\
\hline
\vspace{0.1cm}
$\sigma_{\rm SB}$ (0.3--0.8 keV) & &  $+1.45_{- 1.54} ^{+ 3.33}$ & $-0.35_{- 2.58} ^{+ 2.53}$ & $+2.32_{- 2.86} ^{+ 4.24}$ \\
\vspace{0.1cm}
$\sigma_{\rm SB}$ (0.5--2 keV)   & &  $+1.42_{- 1.58} ^{+ 2.99}$ & $-0.61_{- 2.56} ^{+ 2.60}$ & $+2.18_{- 2.83} ^{+ 3.85}$ \\
\vspace{0.1cm}
$\sigma_{\rm SB}$ (2--10 keV)    & &  $+0.69_{- 1.36} ^{+ 2.76}$ & $+0.64_{- 3.16} ^{+ 2.80}$ & $+0.84_{- 3.16} ^{+ 3.56}$ \\
$\sigma_y$                       & &  $+0.82_{- 1.38} ^{+ 2.72}$ & $+1.23_{- 2.57} ^{+ 3.17}$ & $-0.05_{- 3.90} ^{+ 4.80}$ \\
\hline
No correction                    & &  $+8.26_{- 2.54} ^{+ 4.03}$ & $-2.23_{- 3.59} ^{+ 2.25}$ & $11.15_{- 3.17} ^{+ 6.16}$ \\
\hline
\hline
\label{tab:bestfit_all}
\end{tabular}
\end{center}
\end{table*}

\begin{figure}
\includegraphics[width=0.49\textwidth]{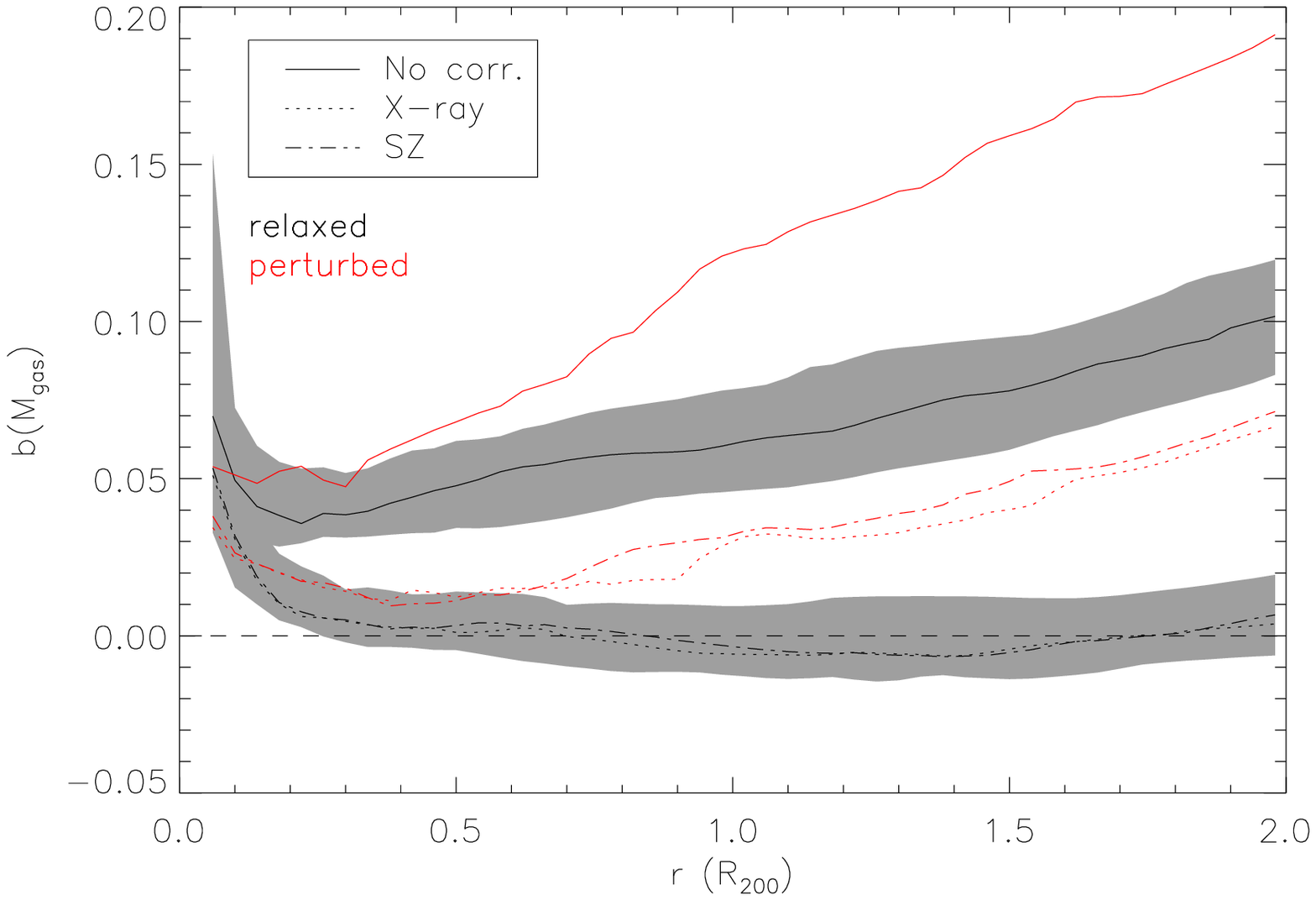}
\caption{Bias in the value of \mgas\ as a function of distance from the centre for 
our reference physical model without correction (solid lines) and with the correction 
obtained by estimating the value of \crhov\ from the scatter in the 2--10 keV surface 
brightness and \ypar\ (dotted and dot-dashed lines, respectively). The black lines 
represent the median of the 30 relaxed objects and the grey-shaded regions enclose the 
quartiles of the uncorrected and of the X-ray corrected biases (the quartile region of 
the SZ corrected bias has a similar size). The red lines represent the median value for 
the perturbed sample. The horizontal dashed line corresponds to the value of 
$b($\mgas$)$=0.}
\label{fig:bias_rad}
\end{figure}

In Fig.~\ref{fig:bias_rad} we show the radial dependence of $b($\mgas$)$ and of the 
corresponding bias after applying the X-ray and SZ corrections. Even close to the core, 
the uncorrected bias is significant ($\sim5$ per cent) and it grows linearly up to 10 
(20) per cent at \rtwoh\ for relaxed (perturbed) systems. We observe that already at 
distances corresponding to $R_{500}$ ($\approx 0.7$\rtwoh) the expected overestimate is 
already close to the values at \rtwoh, with differences of the order of 2 per cent. When 
applying the corrections, we can clearly see that for relaxed objects the systematics is 
completely removed with value a of $0\pm2$ per cent up to 2\rtwoh. Even for perturbed 
haloes the improvement is consistent (a factor of $\sim$3) through the whole radial 
range.

\subsection{Correcting the hydrostatic-equilibrium mass bias}
\label{ssec:corr_mhe}

A frequently used method to determine the total mass of a galaxy cluster relies on the 
assumption that the ICM is in hydrostatic equilibrium. The measured density and 
temperature profiles are then used to determine the total mass $M_{\rm he} (<r)$ enclosed 
by a given radial distance $r$
\begin{equation}
M_{\rm he} (<r) = - \frac{k_{\rm B}T(r) \, r}{G\mu m_{\rm p}} 
                    \left( 
                    \frac{d\log{\rho(r)}}{d\log{r}} + 
                    \frac{d\log{T(r)}}{d\log{r}} 
                    \right) \ ,
\label{eq:mhe}
\end{equation}
where $T(r)$ is the (mass-weighted) temperature profile, and $\mu$ is the mean molecular 
weight in units of the proton mass, $m_{\rm p}$. This method is known to be affected by 
several systematics associated to the break of the spherical symmetry and to the presence 
non-thermal pressure sources that affect particularly cluster outskirts (see 
Section~\ref{sec:intro}). Here, instead, we focus on determining the bias associated to 
the residual clumpiness of the ICM.

It is easy to show that when accounting for the gas density bias of eq.~(\ref{eq:rhox}), 
the X-ray measured hydrostatic-equilibrium mass $M_{\rm he, X}$ formula contains an 
additional term associated to \crhov, thus becoming
\begin{align}
M_{\rm he, X} (<r) = & - \frac{k_{\rm B}T(r) \, r}{G\mu m_{\rm p}} 
                              \left( 
                              \frac{d\log{\rho(r)}}{d\log{r}} + 
                              \frac{d\log{T(r)}}{d\log{r}} + \right. \nonumber \\
                            & + \left. \frac{1}{2}\frac{d\log{C_\mathcal{R}(r)}}{d\log{r}}  
                              \right) \ .
\label{eq:mhe_corr}
\end{align}
Since \crhov$(r)$ generally grows with $r$ (see Figs.~\ref{fig:clp_phys} and 
\ref{fig:clp_samp}), unlike $\rho(r)$ and $T(r)$, this turns into an underestimate of the 
value of \mhe. To study this effect in detail, for every object of our sample we compute 
the true value of \mhe\ and $M_{\rm he, X}$ by adopting the following procedure. We first 
fit $T(r)$ in the range 0.8--1.2 \rtwoh, with the formula of 
\cite{vikhlinin06}\footnote{Here we consider only the 
external part of their profile, i. e. the formula of their eq.~(4), after fixing 
$a=1$ and $b=2$, thus leaving only three free parameters: normalisation, core radius and 
the external slope $c$. However, since our objective is to determine the bias associated 
to the uncertainties in the density profile, any possible effect introduced by changing 
the temperature fitting procedure is negligible.}. Then, in the same radial range, we fit 
$\rho(r)$ and the biased density $\rho_{\rm X}(r)$ profiles with a $\beta$-model 
\citep{cavaliere78} and use these results to compute \mhe\ and $M_{\rm he, X}$, 
respectively, with eq.~(\ref{eq:mhe}) for $r=$\rtwoh.

\begin{figure}
\includegraphics[width=0.49\textwidth]{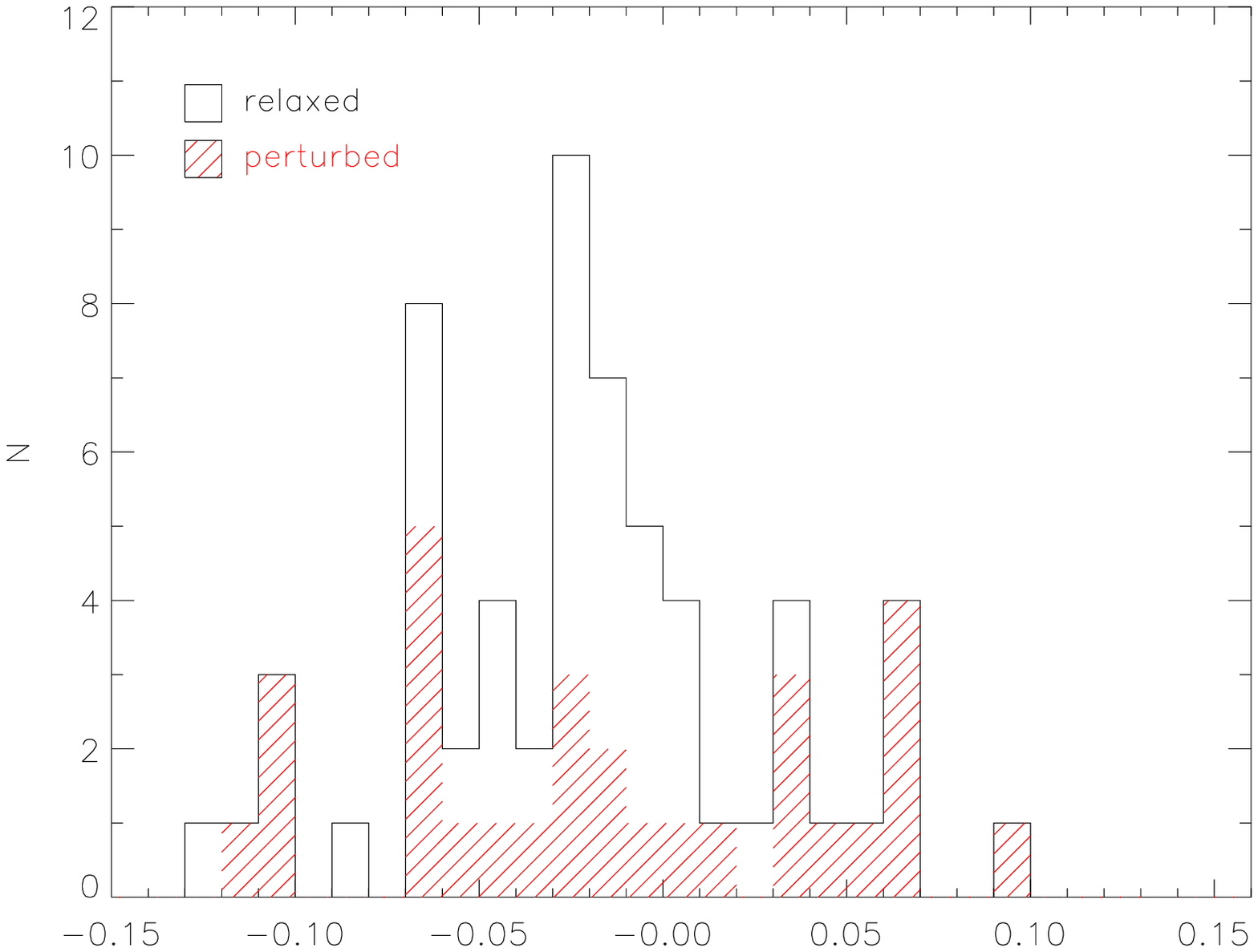}
\includegraphics[width=0.49\textwidth]{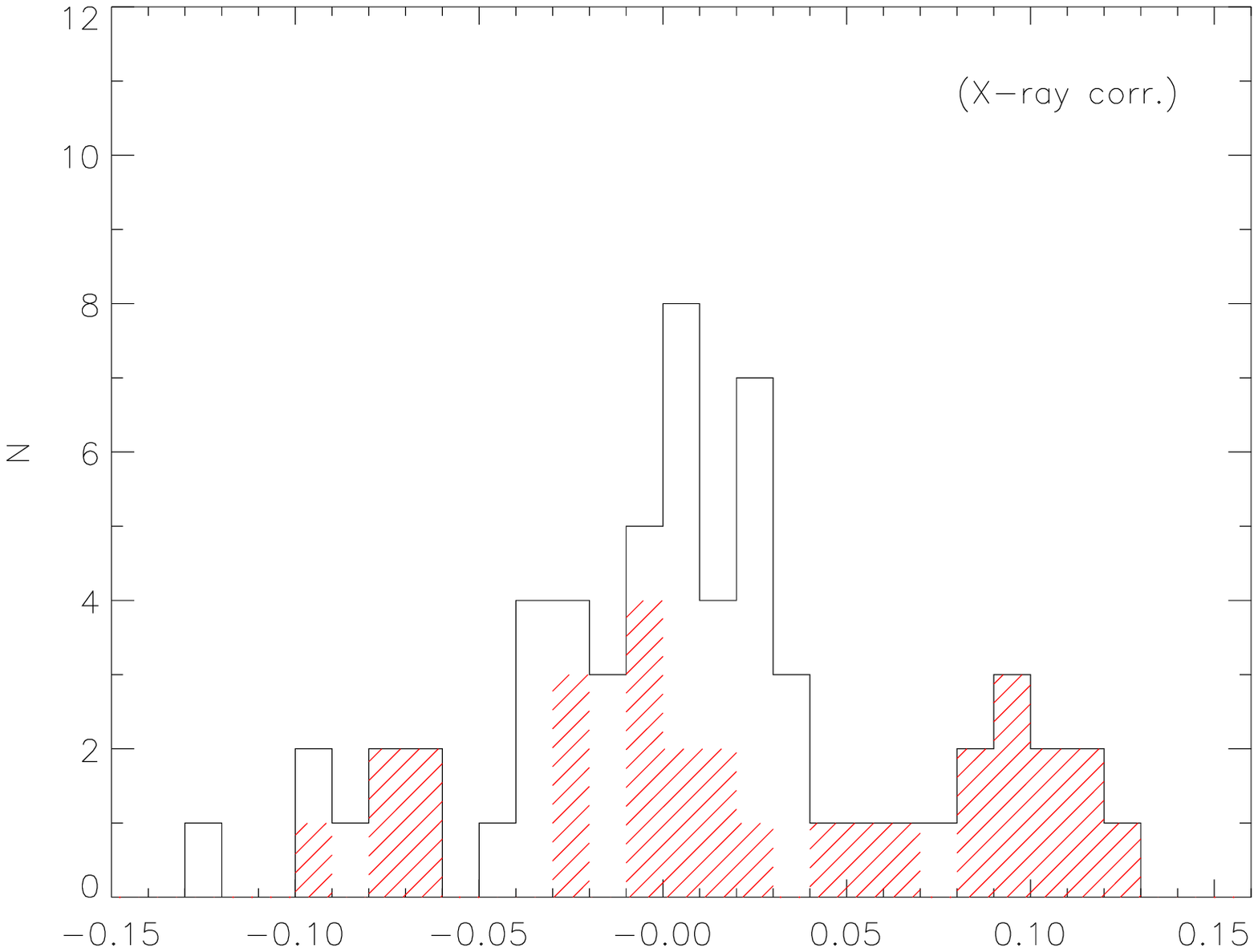}
\includegraphics[width=0.49\textwidth]{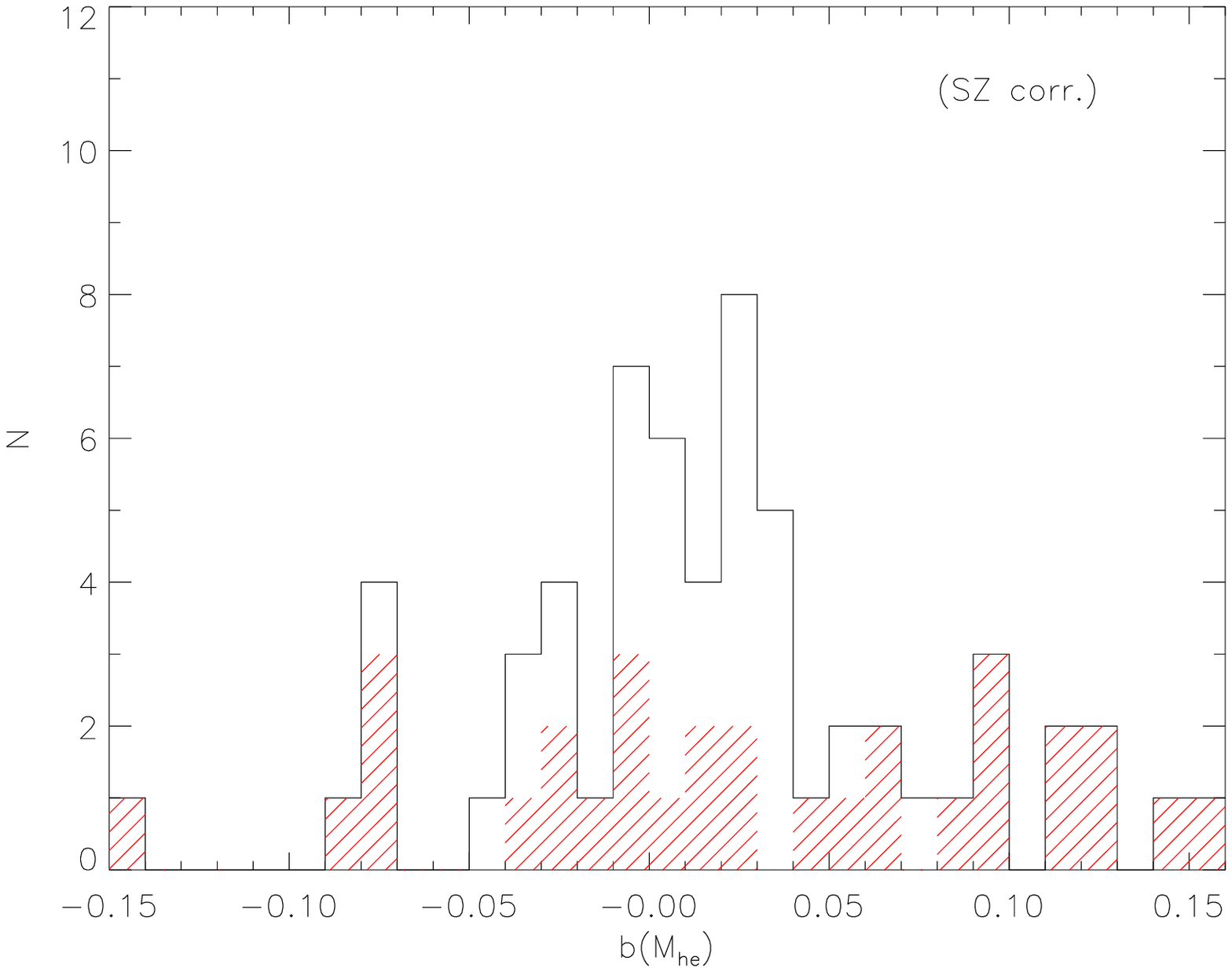}
\caption{Same as Fig.~\ref{fig:bias} but for the bias in the hydrostatic-equilibrium 
mass. One perturbed haloes, D03-\emph{a}, is not shown since its values are outside the 
plot range being, from top to bottom, $b($\mhe$)= -0.22, -0.21$ and $-0.19$, 
respectively. D27-\emph{a} is also outside the plot range in the top and central panels 
with $b($\mhe$)= -0.25$ and $-0.17$, respectively.}
\label{fig:hebias}
\end{figure}

The top panel of Fig.~\ref{fig:hebias} shows the distribution of the expected value of 
$b($\mhe$)$ for our sample of simulated clusters and 
groups. While it is confirmed that for the majority of the cases (45 over 62) the 
residual clumpiness introduces an underestimate, since the clumping factor of 
single objects can be affected by local variations (i.e. \crhov$(r)$ is not perfectly 
monotonic), we end up with about a quarter of our sample in which the measured \mhe\ is 
higher than the true value. As a general remark, the median effect of about $-2$ per cent 
is small when compared to the other systematics, and with respect to the intrinsic 
dispersion ($\sim 6$ per cent difference between the upper and lower quartile). We note, 
however, that for the most disturbed systems the error can reach 10 per cent or more in 
either direction.

We repeat the same procedure done for $b($\mgas$)$ (see Section~\ref{ssec:corr_mgas}) to 
see whether our clumpiness-correction method is effective in reducing the value of 
$b($\mhe$)$. The results for the three X-ray bands and for the SZ effect are shown in 
Table~\ref{tab:bestfit} and \ref{tab:bestfit_all} and in the central and bottom panel of 
Fig.~\ref{fig:hebias} for the hard 2--10 keV band and for the SZ effect only, 
respectively. Also in this case we observe a global improvement of the measurements, with 
the systematics that disappear in all cases. On the other side, the halo to halo scatter 
is not significantly reduced.


\subsection{Correcting the gas fraction bias}
\label{ssec:corr_fgas}

The two biases associated to the residual clumpiness previously described (positive bias 
for \mgas\ and negative for \mhe) add together when measuring the gas fraction \fgas.
In fact, being the measured value
\begin{equation}
f_{\rm gas, X} = \frac{M_{\rm gas, X}}{M_{\rm he, X}} \ ,
\label{eq:fgas}
\end{equation}
the corresponding bias is
\begin{equation}
b(f_{\rm gas}) \equiv \frac{f_{\rm gas,X}-f_{\rm gas}}{f_{\rm gas}} = 
                      \frac{b(M_{\rm gas})-b(M_{\rm he})}{1+b(M_{\rm he})} \ .
\label{eq:bfgas}
\end{equation}

\begin{figure}
\includegraphics[width=0.49\textwidth]{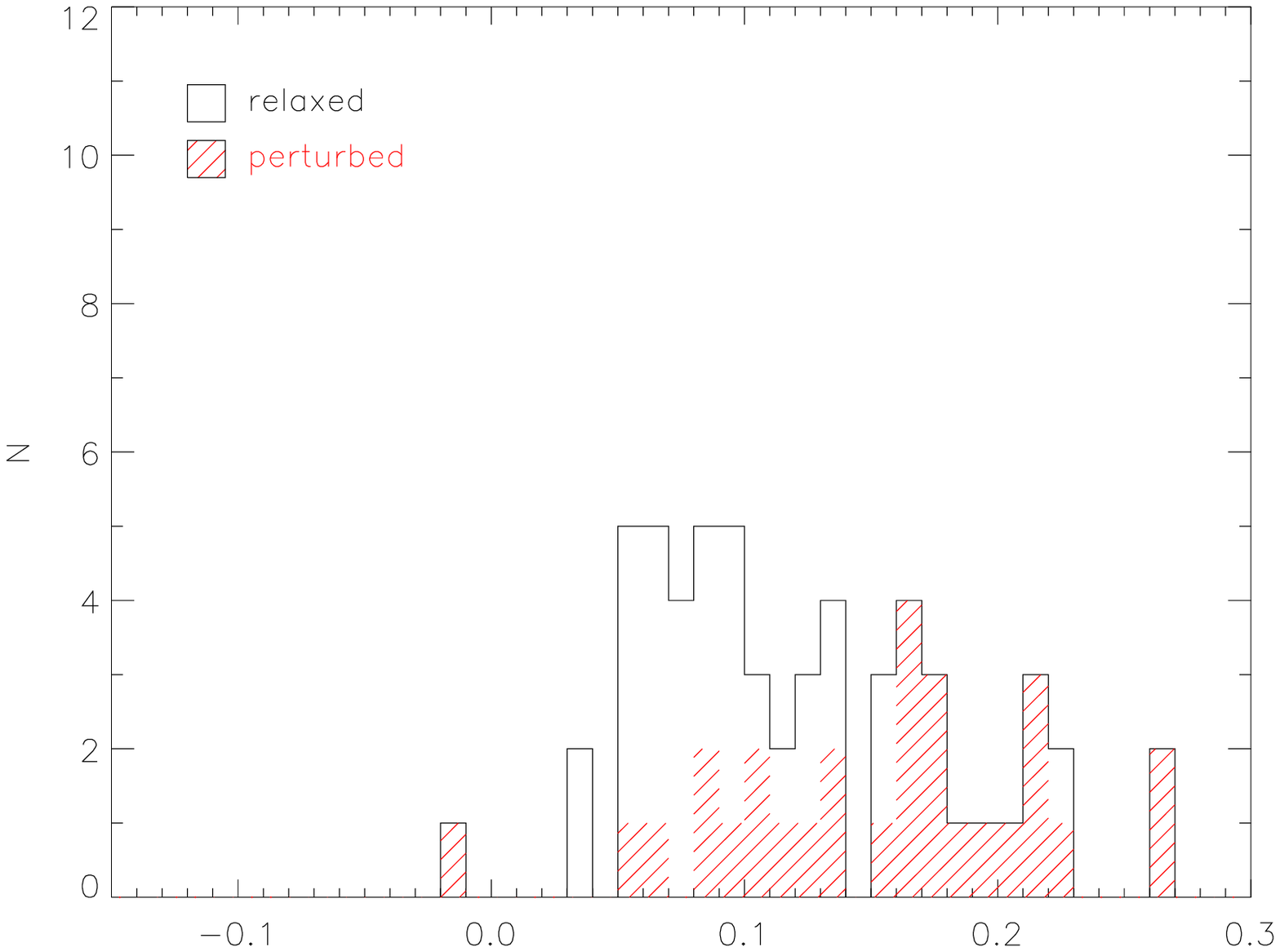}
\includegraphics[width=0.49\textwidth]{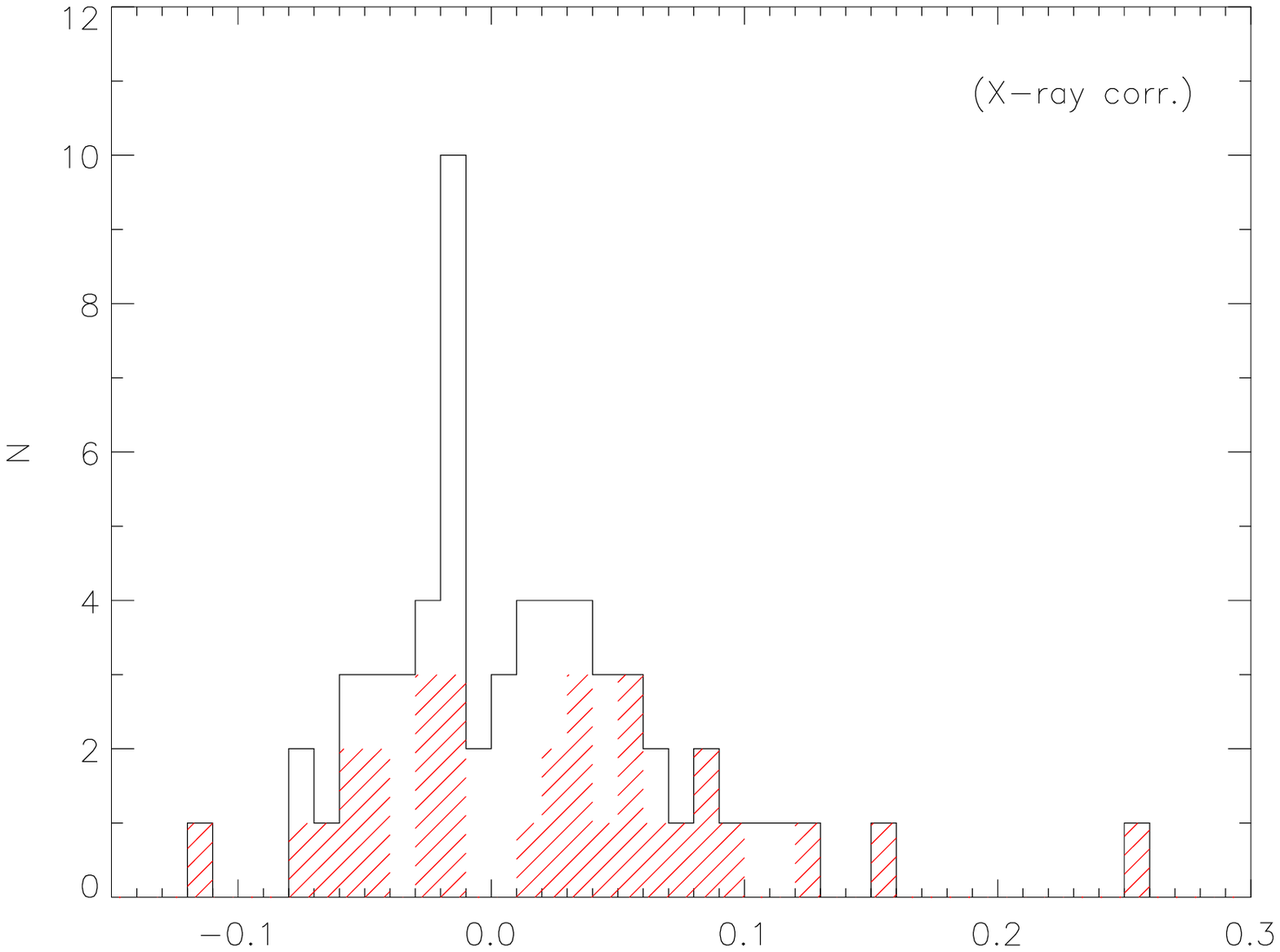}
\includegraphics[width=0.49\textwidth]{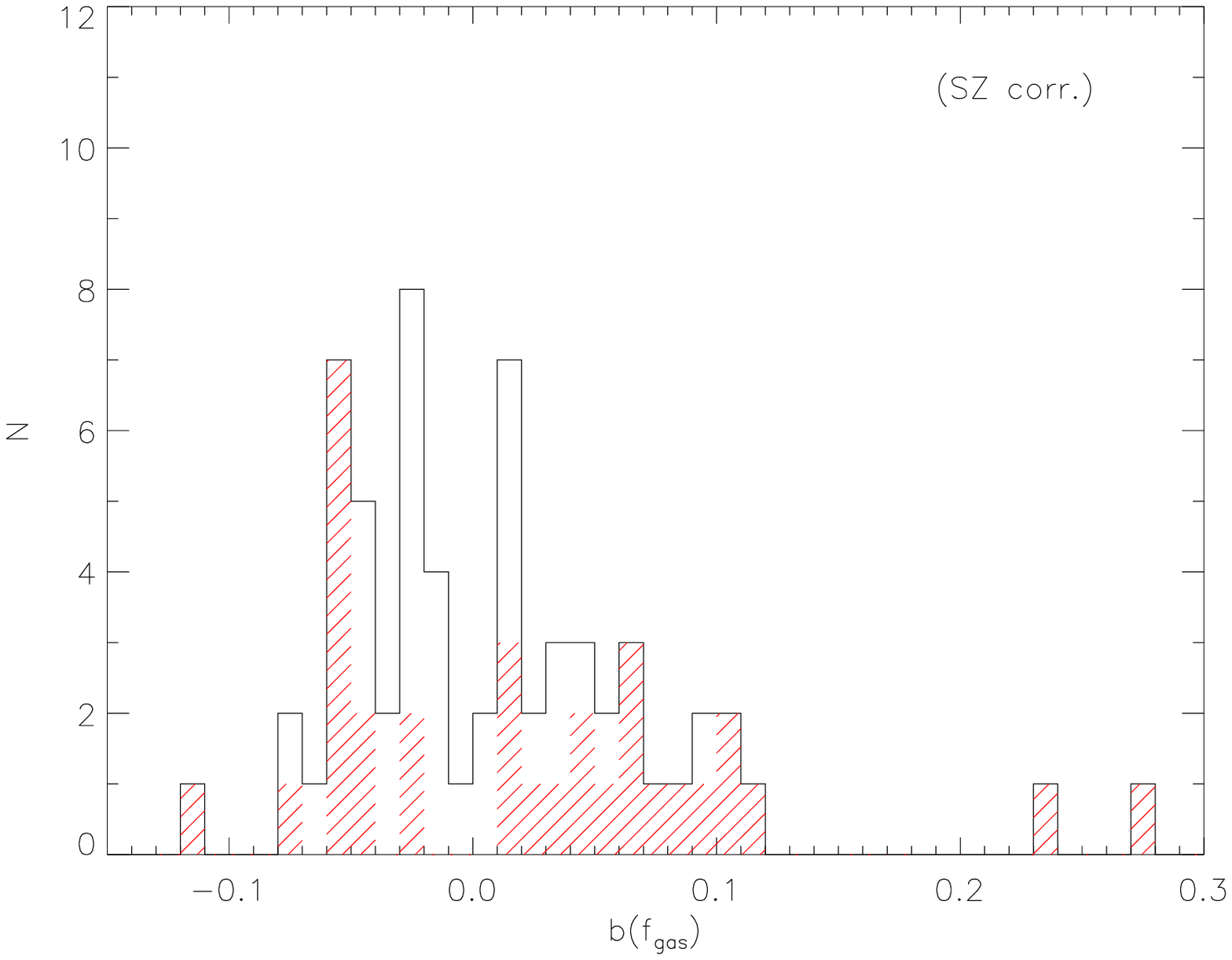}
\caption{Same as Fig.~\ref{fig:bias} but for the bias in the gas fraction. Three perturbed 
haloes, D03-\emph{a}, D18-\emph{a} and D27-\emph{a}, are not shown in the top panel 
since their values are outside the plot range: $b($\fgas$)= 0.35, 0.33$ and $0.65$, 
respectively. D27-\emph{a} is outside the plot range also in the central panel, 
having $b($\fgas$)= 0.30$.}
\label{fig:fgasbias}
\end{figure}

When computing the value of $b($\fgas$)$ at $r=$\rtwoh\ for our sample of 62 haloes, we 
can see from the distribution shown in the top panel of Fig.~\ref{fig:fgasbias} that this 
results in a systematic overestimate on average of about 10 per cent. The ICM 
inhomogeneities cause also a large spread in the measured value, with 11 objects having 
$b($\fgas$)>0.2$ (three are outside the plot range).

We verify the efficiency of our bias-correction method for \fgas\ by computing the 
expected value of $b($\fgas$)$ once the corrected values of $b($\mgas$)$ and $b($\mhe$)$ 
(see Section~\ref{ssec:corr_mgas} and \ref{ssec:corr_mhe}, respectively) are used in 
eq.~(\ref{eq:bfgas}).  We show the results in Table~\ref{tab:bestfit} and 
\ref{tab:bestfit_all} and in the central and bottom panel of Fig.~\ref{fig:fgasbias} 
(2--10 keV band and SZ effect only, respectively). Again, in all cases the average bias 
is significantly reduced, with 
\sigmasb\ for the 2--10 keV band and \sigmay\ that provide the best results. In this 
case, as for $b($\mhe$)$, the application of our method allows also to reduce the scatter 
in the measured values.


\section{Summary and conclusions}\label{sec:concl}
In this paper we have analyzed a set of 62 simulated clusters and groups, obtained with 
different physical prescriptions, with the objective of providing a global 
characterization of their density inhomogeneities in the regions close to the virial 
radius. We have described a method that allows to separate the ICM clumpiness associated 
to small clumps from a ``residual'' one that corresponds to large-scale 
inhomogeneities, and discussed how the latter depends on the mass and the dynamical state 
of the halo. Finally, we have discussed how the presence of large-scale inhomogeneities 
can bias the estimates of \mgas, \mhe\ and \fgas\ and provided a method to reduce this 
bias by using a directly observable quantity: the azimuthal scatter in the X-ray surface 
brightness profiles or in the thermal SZ ones (\ypar\ profiles).

Our results can be summarized as follows.

\begin{itemize}
\item[{\it (i)}]As expected, the degree of global clumpiness in our simulated objects 
depends mainly on the presence/absence of radiative cooling, making about one order of 
magnitude difference. Once cooling is included, additional feedback mechanisms do not 
change significantly the clumping factor.
\item[{\it (ii)}]When considering only the contribution of emitting gas, we obtain 
values of \crho$\simeq 2-3$ at \rtwoh. When compared to the value of $\sim 16$ claimed by 
\cite{simionescu11}, we conclude that their estimate can not be reproduced by our models, 
even neglecting the possibility of identifying emitting clumpy structures.
\item[{\it (iii)}]We introduce the concept of \emph{residual clumpiness}, \crhov, 
to quantify the amount of large-scale inhomogeneities (departure from spherical symmetry, 
presence of filaments), that corresponds to the bulk clumpiness once obvious bright 
condensed clumps are masked out. This quantity is independent of the physical model 
assumed, while it is sensitive to the dynamical state of the halo: relaxed objects 
have \crhov$\sim1.2$ at \rtwoh, while dynamically perturbed ones have on average 
\crhov$\sim 1.5$.
\item[{\it (iv)}]The residual clumpiness of the ICM causes a significant overestimate in 
the measurement of \mgas\ from X-ray observations of the order of $\sim 5-10$ per cent 
for the more relaxed objects up to $\sim 20-30$ per cent for the more perturbed ones.
\item[{\it (v)}] A smaller negative bias of about 2 per cent is present also in the 
measurement of \mhe. Consequently, the combination of the two biases results in an 
overestimate of \fgas, with average values of $\sim 10$ per cent, and an intrinsic 
scatter of $\sim9$ per cent (interquartile range). These biases are lower when 
compared to other known systematics.
\item[{\it (vi)}]The residual clumpiness correlates well ($r_{\rm S} = 0.6-0.7$) 
with the azimuthal scatter of the X-ray surface brightness and of the \ypar\ profiles. This 
allows us to obtain an analytical formula to estimate it as a function of two observable 
variables: the azimuthal scatter and the radial distance.
\item[{\it (vii)}]This relation provides a method to correct the gas density estimates, 
making it possible to improve consistently the accuracy of the \mgas\ measurements. With 
this method the systematics described above disappears completely 
for relaxed haloes from outside the cluster core up 2\rtwoh. For perturbed 
clusters/groups the overestimate is reduced by a factor of about 3.
\item[{\it (viii)}]Finally, this method works also in eliminating the bias associated to 
the measurement of \mhe\ and \fgas. However a large intrinsic scatter (5--7 per cent, 
in terms of interquartile range) between the different objects remains.
\end{itemize}

Overall, our results show how the study of the outskirts of galaxy clusters and groups 
is important for the measurement of the gas mass and gas fraction, and how the 
combination of simulations and observations can improve their precision. A possible 
extension and improvement of this analysis may be investigating the correlation of 
\crhov\ with other inhomogeneities parameters such as the halo ellipticity, or by 
determining how the \crhov$(\sigma,r)$ relations may vary as a function of the observed 
relaxation parameters of the haloes.


\section*{acknowledgments}
Most of the computations necessary for this work have been performed at the Italian 
SuperComputing Resource Allocation (ISCRA) of the Consorzio Interuniversitario del Nord 
Est per il Calcolo Automatico (CINECA). We acknowledge financial contributions from
contracts ASI-INAF I/023/05/0, ASI-INAF I/088/06/0, ASI I/016/07/0 COFIS, ASI Euclid-DUNE 
I/064/08/0, ASIUni Bologna-Astronomy Dept. Euclid-NIS I/039/10/0, ASI-INAF
I/023/12/0, the European Commissions FP7 Marie Curie Initial Training Network CosmoComp 
(PITN-GA-2009-238356), by the PRIN-MIUR09 {\it ``Tracing the growth of structures in the 
Universe''} and by PRIN MIUR 2010-2011 {\it ``The dark Universe and the cosmic evolution 
of baryons: from current surveys to Euclid''}. DF acknowledges funding from the Centre of 
Excellence for Space Sciences and Technologies  SPACE-SI, an operation partly financed by 
the European Union, European Regional Development Fund and Republic of Slovenia, Ministry 
of Education, Science, Culture and Sport. We thank an anonymous referee that provided 
useful indications to improve the quality of our work. We acknowledge useful discussions 
with N.~Cappelluti, G.~Murante, E.~Rasia and L.~Tornatore. We are grateful to D.~Eckert 
for providing us the data on the scatter of his observed profiles.
\bibliographystyle{mn2e}

\newcommand{\noopsort}[1]{}


\appendix
\section{Indentifying clumps with observations}
\label{app:clumps_obs}

Our volume-clipping method, described in Section~\ref{ssec:volume} and more in detail 
in \cite{roncarelli06b}, allows us to separate between the gas belonging to clumps and to 
the bulk of the cluster based on theoretical considerations. Here we provide more 
information on the physical properties of these clumps and investigate up to which extent 
they may be detected in simulated X-ray surface brightness maps, and whether the 
different methods would introduce significant biases in the \crhov\ determination.

\begin{figure}
\includegraphics[width=0.49\textwidth]{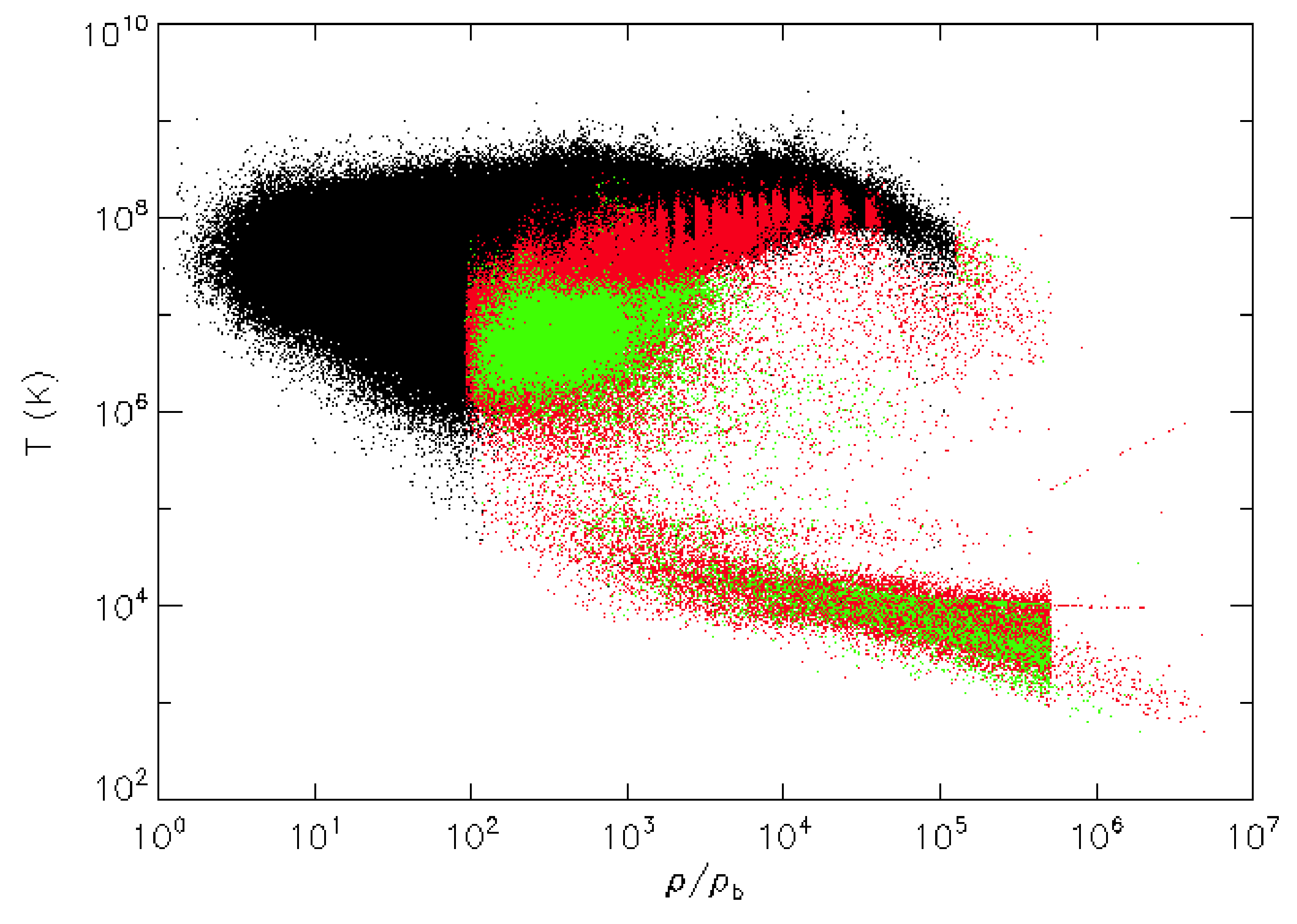}
\caption{Phase diagram of the SPH particles outside 0.5\rtwoh\ of the D17-a simulated 
cluster for our reference model. The black points indicate the particles of the bulk.
The particles identified as clumps by our algorithm are marked in green if they fall 
inside a contour region of the right panel of Fig.~\ref{fig:clump_maps}, and red 
otherwise.}
\label{fig:rhot_clumps}
\end{figure}

We show in Fig.~\ref{fig:rhot_clumps} the $T-\rho$ scatter plot of the SPH 
particles in the outskirts of the D17-a relaxed cluster. The amount of gas that our 
algorithm associates to the clumpy phase (red and green points) corresponds to about the 
3 per cent of the total gas mass inside \rtwoh\ for this halo. We verified that in 
perturbed systems it is slightly higher (4--5 per cent). By analysing the plot, the 
multi-phase nature of these clumps shows up clearly. Part of the gas is associated to the 
cold star-forming phase at $\rho>10^3\rho_{\rm b}$ and $T=10^4-10^5$K. These particles 
are responsible for most of the global ICM clumpiness (see the comments on 
Fig.~\ref{fig:clp_phys}) but would not influence any X-ray measurement since their 
temperature is too low to produce any significant emission. The majority of clump 
particles, which usually embed the previous ones, are instead at higher temperature 
($T> 10^6$K) and can eventually produce a detectable X-ray emission.

\begin{figure*}
\includegraphics{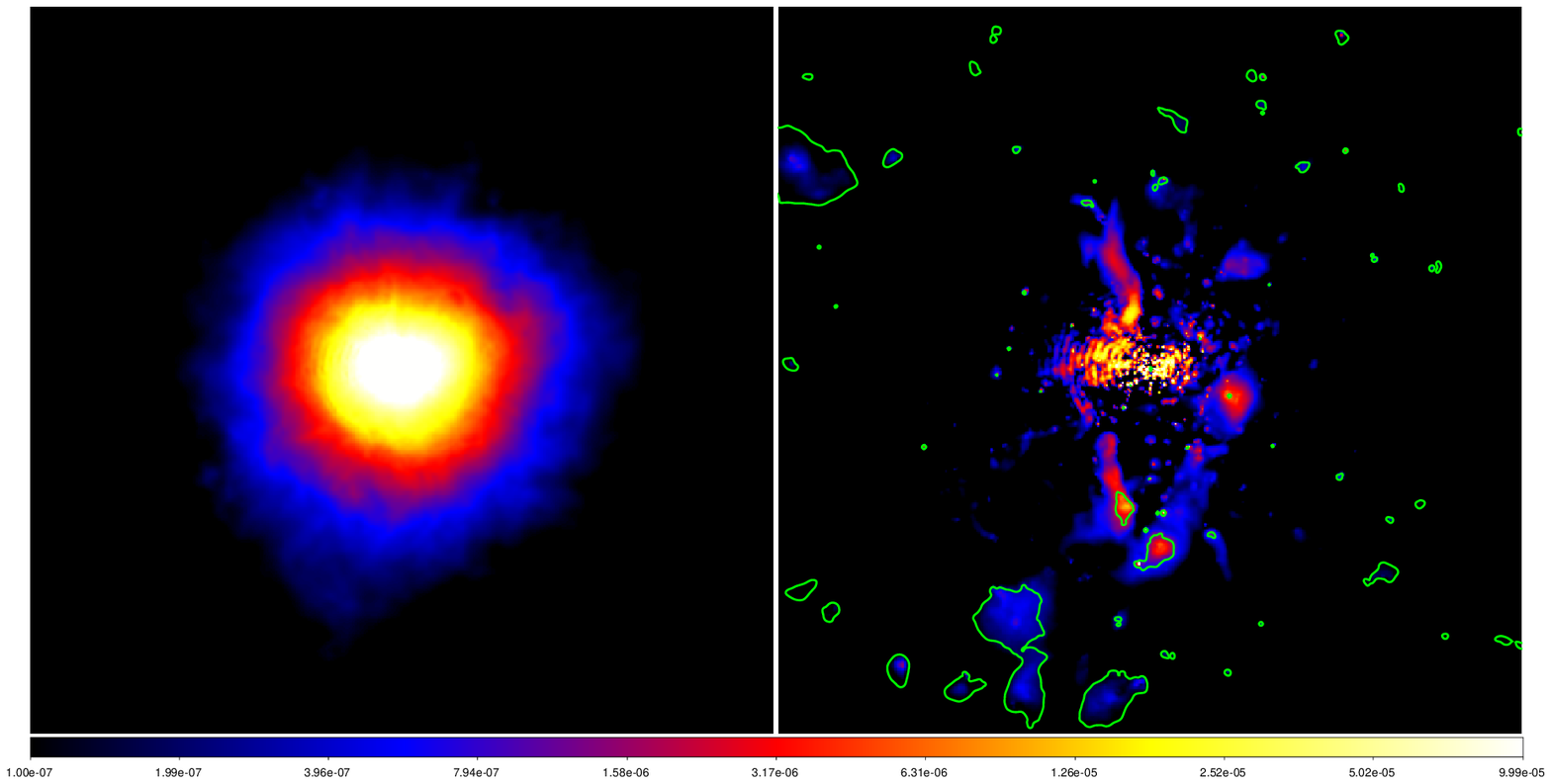}
\caption{Maps of the soft (0.5--2 keV) X-ray surface brightness of the D17-a cluster for 
our reference model in units of counts s\pmone\ cm\pmtwo\ arcmin\pmtwo. Each map is 
centered on the cluster centre and encloses a square of 4\rtwoh\ per side. Left panel: 
the gas which is identified as 'bulk' by our theoretical filtering method (99 per cent 
volume). Right panel: the gas which is identified as 'clumps' (one per cent volume). The 
contours enclose the regions where the projected clumps signal is more than 3 times the 
bulk one, mimicking a S/N$>$3 threshold.}
\label{fig:clump_maps}
\end{figure*}

The maps of Fig.~\ref{fig:clump_maps} show the 0.5--2 keV surface brightness, up to 
2\rtwoh\ of the bulk (left panel) and of the clumps (right) of the D17-a relaxed 
cluster, as defined by our method. The great homogeneity of the bulk map shows clearly 
that no clumpy structure is missed by our filtering method and that any possible 
detectable inhomogeneity must necessarily be associated to the clump phase.
The possibility of detecting them depends on how much their signal is brighter 
with respect to the bulk one. To this purpose we identified the map regions where the 
signal-to-noise ratio exceeds a value of 3, by considering the bulk map as a reference 
for the noise. Most of the clumpy structures are at least partially identified with this 
method, with an increasing efficiency towards the outskirts.

\begin{figure}
\includegraphics[width=0.49\textwidth]{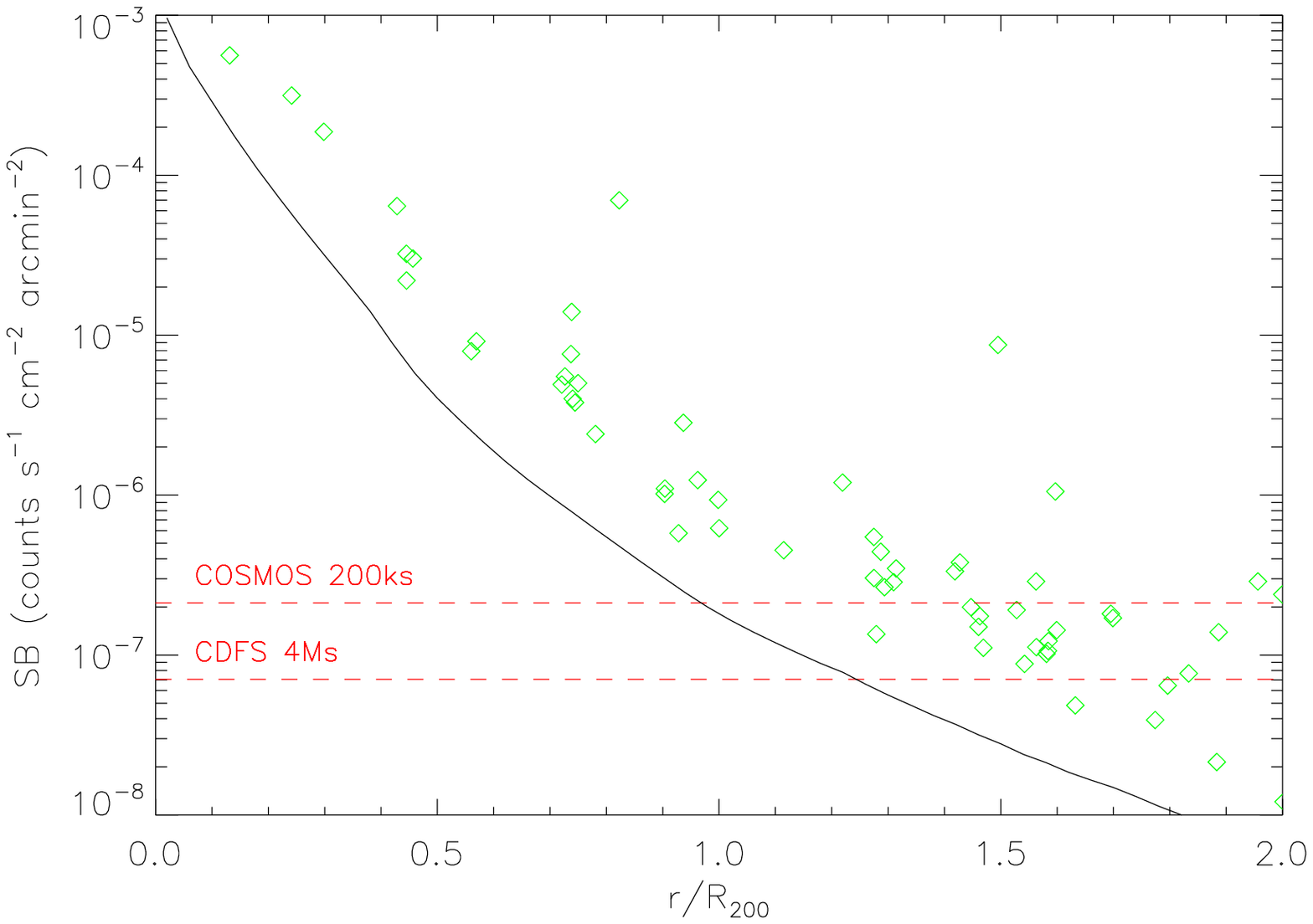}
\caption{Soft (0.5--2 keV) X-ray surface brightness of the clumps (green diamonds) 
identified with the method shown in Fig.~\ref{fig:clump_maps} as a function of distance 
from the centre. The black solid line represents the average surface brightness of the 
cluster bulk. The red dashed lines indicate the measurements of the unresolved X-ray 
background in the same band from the \emph{COSMOS} survey \protect\citep[][ 
$\sim$200ks exposure, upper line]{elvis09} and from the 4Ms \chandra\ Deep Field South 
\protect\citep[][lower line]{cappelluti12}.}
\label{fig:dclumps}
\end{figure}

In Fig.~\ref{fig:dclumps} we show the surface brightness of the identified clumps as a 
function of distance from the centre, compared to the average cluster surface brightness 
profile. Since the central part of these clumps is usually very bright, in most of the 
cases their average surface brightness exceeds the cluster one by more than 5. It is 
clear also, that these structures can be identified also outside \rtwoh, 
since their signal is well above the unresolved X-ray background, at least in the case 
of a massive halo.

As a drawback of our method, when in a given radial bin no clear clumpy structure is 
present we consider as clumps also a small fraction of diffuse gas whose signal is not 
high enough to be identified. This happens in particular in the regions close to the 
centre where the ICM is more uniform, as it can be seen both in the right panel of 
Fig.~\ref{fig:clump_maps} and by the red points of Fig.~\ref{fig:clump_maps}. A precise 
determination of the fraction of clumps that would be missed by real X-ray observations 
clearly depends on instrumental details and is beyond the aim of our work which focuses 
on the large-scale inhomogeneities. However, using our mock maps we can provide an 
estimate of the amount of missed clumpy gas, together with its influence on the bias of 
the gas mass measurements. We verified that the detected clumpy ICM (e.g. green particles 
in Fig.~\ref{fig:clump_maps}) is the $\sim$20 per cent of the total amount of gas 
associated to clumps by our volume-clipping method. In order to have an estimate of its 
impact on the mass gas measurement, we repeated our analysis by reducing by a factor of 
5 the volume threshold (i.e. the 0.2 per cent of the volume of each bin).
We obtained a value of $b($\mgas$)=7.7$ per cent for the relaxed sample, with respect to 
the previous 6.1 (see Table~\ref{tab:bestfit}) obtained with our more conservative 
threshold. Therefore, given the relatively small difference, we conclude that the 
impact on our results of any unresolved clumpy gas is minor with respect to the one 
associated to large-scale inhomogeneities.

\section{Relaxed and perturbed haloes from an observational point of view}

\label{app:rel_ptb}
\begin{figure}
\includegraphics[width=0.49\textwidth]{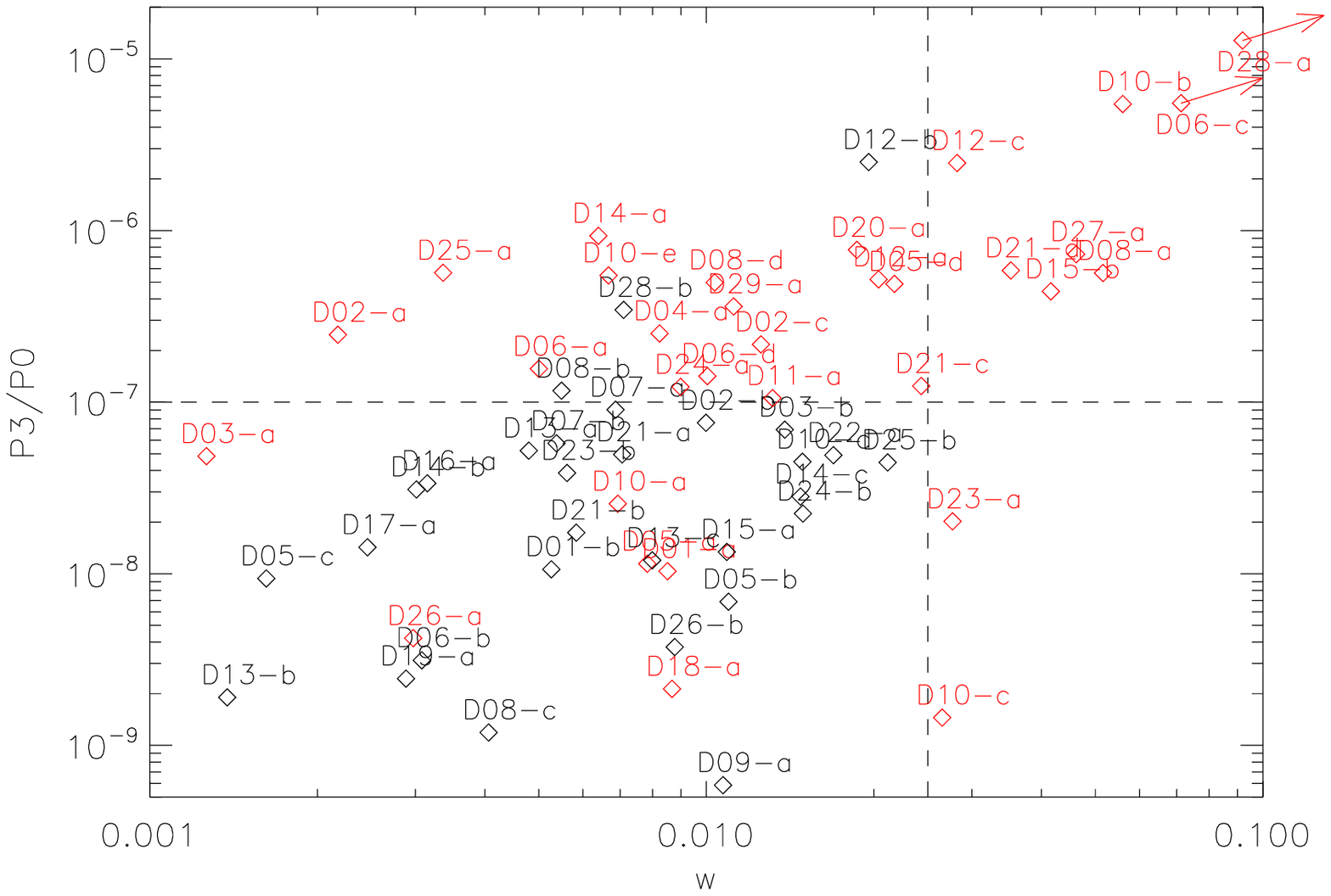}
\caption{Power-ratio $P_3/P_0$ versus the centroid shift $w$ estimated from the 0.5--2 
keV surface brightness maps in the radial range 0.1--1 \rtwoh\ for the sample of our 62 
simulated clusters and groups. The two objects indicated with arrows have values outside 
the plot range. The two dashed lines indicate the limits of $w=0.025$ and 
$P_3/P_0=10^{-7}$ that provide the best match between the two definitions of 
relaxed/perturbed halos (see text).}
\label{fig:rel_diag}
\end{figure}

In Section~\ref{ssec:volume} we have described our classification of simulated haloes 
into \emph{relaxed} and \emph{perturbed} according to the robustness of the determination 
of \crhov. Here we explain how this criterium, which is based on purely theoretical 
considerations, actually matches well with a possible classification done with 
observational methods.

To this purpose we have created a 0.5--2 keV band surface brightness map for each of our 
simulated objects. Following \cite{cassano10}, we have then computed, over the radial 
range $0.1-1 R_{200}$, the centroid shift $w$, defined as the standard deviation, in 
units of \rtwoh, of the projected separation between the X-ray peak and the centroid, and 
the power-ratio $P_3/P_0$ \citep[see][]{buote95}, that is the lowest normalized moment of 
the X-ray surface brightness clearly connected to substructures \citep[see, e.g.,][]
{bohringer10}. We show in Fig.~\ref{fig:rel_diag} the position of our simulated systems 
in the $P_3/P_0$ vs $w$ plane. Haloes defined as relaxed according to our definition 
(marked in black) have on average lower values of all the parameters, thus sitting on the 
lower left corner of the plot, while perturbed ones tend to show a high value in at least 
one of the two parameters.

The dashed lines ($w=0.025$ and $P_3/P_0=10^{-7}$) show the limits that roughly 
correspond to the best match between the two classifications. When defining 
observationally the relaxed haloes as the ones laying on the bottom-left quadrant of the 
plot, and perturbed otherwise, we end up with 53 over 62 haloes matching the 
corresponding theoretical definition.


\label{lastpage}
\end{document}